\documentclass[aps,pra,showpacs,floatfix,longbibliography,superscriptaddress,nofootinbib]{revtex4-2}
\usepackage{graphicx,amsfonts,amssymb,amsmath}
\usepackage{bm,booktabs,svg}
\usepackage{textcomp} 
\usepackage{esint}
\usepackage[english]{babel}
\babelprovide{en}

\usepackage{afterpage}
\usepackage{bbold}
\usepackage{soul}
\usepackage{svg} % for vectorial figure
\usepackage[colorlinks,linktocpage,citecolor=blue,linkcolor=red,urlcolor=blue]{hyperref}
\usepackage[dvipsnames]{xcolor}     
\usepackage[todonotes={textsize=scriptsize}]{changes}
\definechangesauthor[name=Yaniss, color=MidnightBlue]{YR}
\definechangesauthor[name=Adam, color=ForestGreen]{AR}

\begin{document}
    
\graphicspath{{figures_submit/}}
	
\allowdisplaybreaks
	
% texdef -s \j

% Livre avec corrections en rouge et remarques
\newcommand{\oldnew}[2]{\marginpar{\scriptsize \textcolor{red}{correction}}{\textcolor{red}{#2}}}
\newcommand{\suppressed}[1]{\marginpar{\scriptsize \textcolor{red}{correction}}{\textcolor{red}{\st{#1}}}}
\newcommand{\correction}[1]{\marginpar{\textcolor{red}{\scriptsize #1}}}
\newcommand{\marge}[1]{\marginpar{\scriptsize #1}}
\newcommand{\remarque}[1]{\marginpar{\scriptsize Remarque}{\it [#1]}}

% Livre avec corrections en noir sans remarques 
%
%\newcommand{\oldnew}[2]{#2}
%\newcommand{\suppressed}[1]{}
%\newcommand{\marge}[1]{}
%\newcommand{\correction}[1]{}
%\newcommand{\remarque}[1]{}

%%%%%%%%%%%%%%%%%%%%%%%%%%%%%%%%%%%%%%%%%%%%%%%%%%%%%%%%%%%%
\def\rhoeq{\hat\rho_{\rm eq}}

\newcommand{\new}[1]{{\bf #1}}
\newlength{\textlarg}
\newcommand{\redbar}[1]{\textcolor{red}{\st{#1}}} % requires package soul 
\newcommand{\bluebar}[1]{\textcolor{blue}{\st{#1}}} % requires package soul

\newcommand{\normord}[1]{:\mathrel{#1}:}

\newcommand{\beq}{\begin{equation}}
\newcommand{\eeq}{\end{equation}}
\newcommand{\bfig}{\begin{figure}}
\newcommand{\efig}{\end{figure}}
\newcommand{\bline}{\begin{multline}}
\newcommand{\eline}{\end{multline}}
\newcommand{\bremark}{\begin{quotation} \noindent \small }
\newcommand{\eremark}{\end{quotation}}
\newcommand{\llbrace}{\left\lbrace}  
\newcommand{\rrbrace}{\right\rbrace}
\newcommand{\lbraket}{\left[}
\newcommand{\rbraket}{\right]}
\newcommand{\llangle}{\left\langle}
\newcommand{\rrangle}{\right\rangle} 

\newcommand{\Tr}{{\rm Tr}} 
\newcommand{\tr}{{\rm tr}} 
\newcommand{\sgn}{\,{\rm sgn}} 
\newcommand{\mean}[1]{\langle #1 \rangle}
\newcommand{\commu}[2]{[#1,#2]} 
\newcommand{\bra}[1]{\langle#1|}
\newcommand{\ket}[1]{|#1\rangle}
\newcommand{\braket}[2]{\langle #1|#2\rangle}
\newcommand{\ketbra}[2]{|#1\rangle\langle#2|}
\newcommand{\dbraket}[3]{\langle #1|#2|#3\rangle}
\newcommand{\tens}[1]{\overleftrightarrow{#1}}  
\newcommand{\vac}{|{\rm vac}\rangle} 
\newcommand{\bravac}{\langle{\rm vac}|}
\newcommand{\const}{{\rm const}} 
\newcommand{\unif}{{\rm unif.}} 
\newcommand{\atanh}{\,{\rm atanh}}
\newcommand{\cotanh}{\,{\rm cotanh}}

\newcommand{\ie}{i.e.\xspace}
\newcommand{\iet}{i.e.}
\newcommand{\eg}{e.g.\xspace}
\newcommand{\cc}{{\rm c.c.}} 
\newcommand{\hc}{{\rm H.c.}} 
\newcommand{\etal}{{\it et al. }}
\newcommand\eme{$^{\mbox{\footnotesize ème}}$\xspace}

\newcommand{\jhatbf}{\hat {\textbf \jold}} 
\newcommand{\Jhatbf}{\hat {\textbf \J}} 
\newcommand{\jhat}{\hat {\jmath}} 
\newcommand{\Jhat}{\hat {J}} 
\newcommand{\jbf}{\textbf j}
\newcommand{\Jbf}{\textbf J}

\def\chibf{\boldsymbol{\chi}}
\def\down{\downarrow}
\def\eps{\epsilon}
\def\gam{\gamma} 
\def\alphabf{\boldsymbol{\alpha}}
\def\phibf{\boldsymbol{\phi}}
\def\varphibf{\boldsymbol{\varphi}}
\def\varphibfs{\boldsymbol{\varphi}_<}
\def\varphibfl{\boldsymbol{\varphi}_>}
\def\varphis{\varphi_{<}}
\def\varphil{\varphi_{>}}
\def\psibf{\boldsymbol{\psi}}
\def\thetabf{\boldsymbol{\theta}}
\def\Ome{\Omega}
\def\omeD{{\omega_D}} 
\def\bfOme{\boldsymbol{\Omega}} 
\def\Omebf{\boldsymbol{\Omega}} 
\def\lamb{\lambda}
\def\Lamb{\Lambda}
\def\sig{\sigma}
\def\Sig{\Sigma}
\def\sigp{{\sigma'}} 
\def\bfsig{\boldsymbol{\sigma}} 
\def\sigbf{\boldsymbol{\sigma}} 
\def\bfSig{\boldsymbol{\Sigma}} 
\def\The{\Theta} 
\def\up{\uparrow}

\def\epsk{\epsilon_{\bf k}} 
\def\epsp{\epsilon_{\bf p}} 
\def\xik{\xi_{\bf k}} 
\def\txik{\tilde\xi_{\bf k}} 
\def\xip{\xi_{\bf p}} 
\def\epsq{\epsilon_{\bf q}} 
\def\xiq{\xi_{\bf q}} 
\def\xikq{\xi_{{\bf k}+{\bf q}}} 
\def\Ek{E_{\bf k}} 
\def\Ep{E_{\bf p}}
\def\Eq{E_{\bf q}}
\def\Heff{\hat H_{\rm eff}}
\def\Hem{\hat H_{\rm em}}
\def\Hint{\hat H_{\rm int}}
\def\Hloc{\hat H_{\rm loc}}
\def\HMF{\hat H_{\rm MF}}
\def\HLL{\hat H_{\rm LL}}
\def\Hdis{\hat H_{\rm dis}}
\def\Sem{S_{\rm em}}
\def\SMF{S_{\rm MF}} 
\def\SHF{S_{\rm HF}} 
\def\SRPA{S_{\rm RPA}} 
\def\Sint{S_{\rm int}} 
\def\Sloc{S_{\rm loc}}
\def\TN{T_{\rm N}} 
\def\TNHF{T^{\rm HF}_{\rm N}} 
\def\Zloc{Z_{\rm loc}} 
\def\ZMF{Z_{\rm MF}} 
\def\ZHF{Z_{\rm HF}} 
\def\ZRPA{Z_{\rm RPA}} 
\def\RPA{{\rm RPA}}
\def\loc{{\rm loc}} 
\def\pp{{\rm pp}}
\def\ph{{\rm ph}} 
\def\ch{{\rm ch}}
\def\sp{{\rm sp}} 
\def\qtf{q_{\rm TF}}
\def\epstf{\eps^{}_{\rm TF}} 
\def\epsrpa{\eps^{}_{\rm RPA}} 
\def\chinnzpp{\chi_{nn}^{0}{}\!\!\!''}
\def\SigHF{\Sigma_{\rm HF}}
\def\psicl{\psi_{\rm cl}} 

\def\half{\frac{1}{2}}
\def\dhalf{\dfrac{1}{2}}
\def\third{\frac{1}{3}} 
\def\quarter{\frac{1}{4}}

\def\qr{{\bf q}\cdot{\bf r}}
\def\wt{\omega t} 

\def\a{{\bf a}}
\def\b{{\bf b}}
\newcommand{\cv}{{\bf c}} 
\def\e{{\bf e}}
\def\f{{\bf f}}
\def\g{{\bf g}}
\def\h{{\bf h}}
\def\jold{\char"11}
\def\j{{\bf j}}
\def\k{{\bf k}}
\def\l{{\bf l}}
\def\ellbf{\bm{\ell}} 
\def\m{{\bf m}}
\def\n{{\bf n}} 
\def\p{{\bf p}} 
\def\q{{\bf q}}
\def\r{{\bf r}}
\def\t{{\bf t}}
\def\u{{\bf u}}
\newcommand{\vv}{{\bf v}}
\def\x{{\bf x}}
\def\y{{\bf y}} 
\def\z{{\bf z}} 
\def\A{{\bf A}}
\def\B{{\bf B}}
\def\D{{\bf D}} 
\def\E{{\bf E}} 
\def\F{{\bf F}} 
\def\H{{\bf H}}  
\def\J{{\bf J}}
\def\K{{\bf K}} 

\def\G{{\bf G}}
\def\L{{\bf L}}
\def\M{{\bf M}}  
\def\O{{\bf O}} 
\def\P{{\bf P}} 
\def\Q{{\bf Q}} 
\def\R{{\bf R}}
\def\S{{\bf S}}
\def\U{{\bf U}} 
\def\V{{\bf V}} 
\def\X{{\bf X}} 
\def\Y{{\bf Y}} 
\def\epsbf{\boldsymbol{\epsilon}}
\def\betabf{\boldsymbol{\beta}}
\def\deltabf{\boldsymbol{\delta}}
\def\mubf{\boldsymbol{\mu}}
\def\nablabf{\boldsymbol{\nabla}}
\def\rhobf{\boldsymbol{\rho}}
\def\sigmabf{\boldsymbol{\sigma}} 
\def\Pibf{\boldsymbol{\Pi}}
\def\pibf{\boldsymbol{\pi}}

\def\para{\parallel}
\def\kpara{{k_\parallel}}
\def\kperp{{k_\perp}} 
\def\kperpp{{k_\perp'}} 
\def\qperp{{q_\perp}} 
\def\tperp{{t_\perp}} 

\def\w{\omega}
\def\wn{\omega_n}
\def\wm{\omega_m}
\def\wnu{\omega_\nu}
\def\wp{\omega_p} 
\def\dmu{{\partial_\mu}}
\def\dnu{{\partial_\nu}}
\def\dl{{\partial_l}}  
\def\dt{\partial_t} 
\def\tdt{\tilde\partial_t}
\def\dk{\partial_k}
\def\tdk{\tilde\partial_k}
\def\dx{\partial_x}
\def\dy{\partial_y} 
\def\dw{\partial_{\w}}
\def\dtau{{\partial_\tau}}  
\def\det{{\rm det}} 
\def\Pf{{\rm Pf}}
\def\diag{{\rm diag}}

\def\dsum{\displaystyle \sum}
\def\dint{\displaystyle \int} 
\def\intt{\int_{-\infty}^\infty dt} 
\def\inttp{\int_{-\infty}^\infty dt'} 
\def\intk{\int_{\bf k}} 
\def\intkd{\int \frac{d^dk}{(2\pi)^d}}
\def\intq{\int_{\bf q}} 
\def\intr{\int d^dr}  
\def\dintr{\displaystyle \int d^dr} 
\def\intrp{\int d^dr'}
\def\dinttau{\displaystyle \int_0^\beta d\tau}
\def\dinttaup{\displaystyle \int_0^\beta d\tau'}
\def\inttau{\int_0^\beta d\tau}
\def\inttaup{\int_0^\beta d\tau'}
\def\intx{\int d^{d+1}x} 
\def\inttaur{\int_0^\beta d\tau \int d^dr}
\def\intinf{\int_{-\infty}^\infty}
\def\dinttaur{\displaystyle \int_0^\beta d\tau \int d^dr}
\def\dintinf{\displaystyle \int_{-\infty}^\infty}
\def\intw{\int_{-\infty}^\infty \frac{d\w}{2\pi}}
\def\sumr{\sum_{\bf r}} 

\def\calA{{\cal A}}
\def\calAbf{\bm{{\cal A}}}
\def\calB{{\cal B}} 
\def\calC{{\cal C}} 
\def\dt{\partial_t}
\def\calD{{\cal D}}
\def\calE{{\cal E}}
\def\calF{{\cal F}} 
\def\calFbf{\bm{{\cal F}}}
\def\calG{{\cal G}}
\def\calH{{\cal H}}
\def\calI{{\cal I}}
\def\calJ{{\cal J}}
\def\calK{{\cal K}}
\def\calL{{\cal L}} 
\def\calM{{\cal M}} 
\def\calN{{\cal N}}
\def\calO{{\cal O}}
\def\calP{{\cal P}}  
\def\calR{{\cal R}} 
\def\calS{{\cal S}}
\def\calT{{\cal T}}
\def\calU{{\cal U}}
\def\calV{{\cal V}}
\def\calX{{\cal X}} 
\def\calY{{\cal Y}} 
\def\calW{{\cal W}} 
\def\calZ{{\cal Z}}

\def\tT{{\tilde T}}
\def\talpha{{\tilde\alpha}}
\def\tbeta{{\tilde\beta}}
\def\tchi{{\tilde\chi}}
\def\tdelta{{\tilde\delta}}
\def\tDelta{{\tilde\Delta}}
\def\teta{{\tilde\eta}} 
\def\tlamb{{\tilde\lambda}}
\def\tmu{{\tilde\mu}}
\def\tphibf{{\tilde\phibf}}
\def\trho{{\tilde\rho}}
\def\tvarphibf{{\tilde\varphibf}} 
\def\tq{\tilde q}
\def\tw{{\tilde\omega}}
\def\twn{{\tilde\omega_n}}
\def\twnu{{\tilde\omega_\nu}}

\def\asinh{{\rm asinh}} 
\def\Tbkt{T_{\rm BKT}}

\title{Universal thermodynamics and BKT transition of the planar Bose gas from weak to strong interactions}

\author{Yaniss Rabahi } 
\affiliation{Univ. Lille, CNRS, UMR 8523 – PhLAM – Laboratoire de Physique des Lasers Atomes et Molécules, F-59000 Lille, France}	

\author{Ivan Balog}
 \affiliation{Institute of Physics, Bijeni\v{c}ka cesta 46, HR-10001 Zagreb, Croatia}
    	
\author{Adam Ran\c{c}on} 
\affiliation{Univ. Lille, CNRS, UMR 8523 – PhLAM – Laboratoire de Physique des Lasers Atomes et Molécules, F-59000 Lille, France}	
 \affiliation{Institut Universitaire de France}

\begin{abstract}
We compute the universal thermodynamics of the dilute two-dimensional Bose gas at weak and strong coupling using the functional renormalization group (FRG). We employ an approximation that retains the full field dependence of the effective potential and spatial derivative terms, extending previous FRG calculations. We find that the equation of state is remarkably insensitive to the approximation scheme, with differences of only a few percent even at strong coupling. By contrast, the improved approximation substantially improves the description of the Berezinskii–Kosterlitz–Thouless (BKT) transition, reproducing the universal jump of the superfluid density and the critical anomalous dimension to within a few percent. At weak coupling, our results agree very well with classical-field Monte Carlo simulations, while at stronger coupling we observe deviations from classical-field scaling, valid at weak coupling. Our approach thus provides a quantitative description of the thermodynamics and superfluid density across the BKT transition, including in the strong-coupling regime where classical-field methods are no longer reliable.
\end{abstract}

\maketitle

\section{Introduction}

Dilute two-dimensional (2D) Bose gases provide a setting to study the effects of strong thermal and quantum fluctuations in interacting systems \cite{popov_1983,fisher_1988,hadzibabic_2011}. 
Their thermodynamics is governed by a quantum critical point at zero temperature and chemical potential \cite{Sachdev_book,ranconUniversalThermodynamicsTwodimensional2012}, corresponding to a vacuum-to-superfluid transition. While this transition happens at zero temperature, it governs the properties of the gas to relatively high temperatures and gives rise to universal scaling properties. In particular, the equation of state depends on the chemical potential and temperature through the ratio $\mu/T$ and a renormalized, energy-dependent interaction \cite{ranconUniversalThermodynamicsTwodimensional2012}.
At finite temperature, the Hohenberg–Mermin–Wagner theorem forbids true long-range order, but a superfluid phase survives below the Berezinskii–Kosterlitz–Thouless (BKT) transition \cite{Hohenberg67,Kosterlitz73,kosterlitz_1974,Berezinskii71,Berezinskii72}. 

Capturing the thermodynamics of an interacting 2D Bose gas is notoriously challenging. The BKT transition itself is of infinite order, the thermodynamic potential exhibits an essential singularity, while the superfluid density displays a universal jump \cite{hadzibabic_2011,nelson_1977}. There is no conventional symmetry breaking, and near the critical point, the system displays a broad fluctuation regime with universal scaling behavior. 
Mean-field approaches fail to describe this regime, as they cannot account for the binding and unbinding of vortex pairs or the infrared divergence of phase fluctuations. As a result, precise theoretical predictions for quantities like the pressure, density, and compressibility in two dimensions require non-perturbative methods that properly include the effects of strong phase fluctuations and the marginal (logarithmic) scale dependence of interactions in 2D. 

Ultracold atomic gases have provided experimental realizations of this physics, allowing for direct observations of the BKT transition \cite{Hadzibabic06,Desbuquois12} and measurements of the equation of state  \cite{Yefsah11,Hung11,Zhang11,Zhang12,haStronglyInteractingTwoDimensional2013,Desbuquois14}. On the theoretical side, classical-field Monte Carlo simulations have established precise results for the weakly interacting 2D Bose gas~\cite{Prokofev01,Prokofev02}. In particular, they determined the critical density and chemical potential, as well as the universal equation of state in the fluctuation regime. These results have been widely used to interpret experiments and provide a benchmark for other theoretical approaches. However, the mapping to a classical field theory relies on weak interactions and becomes inaccurate at stronger coupling. Deviations from classical field theory predictions have already been observed in measurements of the two-body contact at relatively weak interactions \cite{Zou21}.

The functional renormalization group (FRG) \cite{berges_2002,dupuis_2021} offers an alternative approach, which does not rely on a weak-coupling expansion and can describe quantum and thermal fluctuations within the same framework. In Ref.~\cite{ranconUniversalThermodynamicsTwodimensional2012}, the FRG was used to compute the universal equation of state of the dilute 2D Bose gas, obtaining good agreement with experimental measurements. More recently, the method was shown to reproduce measurements of the two-body contact \cite{Rancon23}, including in a regime where classical-field simulations fail. These calculations were based on a relatively simple approximation of the effective action, in which the effective potential was expanded around its minimum and the derivative terms were taken to be field independent. Although sufficient to describe the equation of state, this approximation does not correctly reproduce the low-temperature BKT phase. In particular, the superfluid density always vanishes, preventing a reliable determination of its universal jump.

In this work, we extend the previous FRG approach by retaining the full field dependence of the effective potential and of the spatial derivative terms. This approximation allows us to compute both the equation of state and the superfluid density across the BKT transition, at weak and strong coupling. By comparing the results with those of the simpler approximation, we find that the equation of state is remarkably insensitive to the truncation, with differences of only a few percent even at strong coupling. In contrast, the improved approximation is necessary to better capture the BKT transition, reproducing the universal jump of the superfluid density and the critical anomalous dimension to within a few percent.

At weak coupling, we compare our results with classical-field Monte Carlo simulations~\cite{Prokofev01,Prokofev02}. We obtain very good agreement for the critical parameters and the universal equation of state in the fluctuation regime. At stronger coupling, we observe deviations from classical-field scaling, while the more general quantum scaling form remains valid. Our results therefore provide predictions for the thermodynamics and superfluid density of planar Bose gases beyond the weak-coupling regime, where classical-field methods are no longer quantitatively reliable. This provides a comprehensive and quantitatively reliable picture of the universal thermodynamic behavior of two-dimensional Bose gases, bridging the gap between the weakly and strongly interacting regimes and offering a valuable point of comparison for future experiments and simulations.

The manuscript is organized as follows. In Sec.~\ref{sec_univ}, we recall the main thermodynamic properties of the dilute planar Bose gas and their universal scaling forms. Section~\ref{sec_FRG} presents the FRG formalism and the approximation scheme. We discuss the convergence of the method in Sec.~\ref{sec:convergence_lpade2}, the universal equation of state in Sec.~\ref{sec:equation_etat_universelle}, and the BKT transition and superfluid density in Sec.~\ref{sec:FRGandBKT}. In Sec.~\ref{sec_MC}, we compare our results to classical field simulations in the weak-coupling limit. Our conclusions are given in Sec.~\ref{sec_concl}.

\section{Universal thermodynamics}
\label{sec_univ}
\label{sec:chap3RGWilson}

The physics of a dilute 2D Bose gas at low temperature is described by the Hamiltonian (we use units such that $\hbar=k_B=1$)
\begin{equation}
\begin{split}
    \hat{H} = \int d^2 r \bigg\{
    &
\hat{\psi}^\dagger(\r) \left( -\frac{\nabla^2}{2m} - \mu \right) \hat{\psi}(\r)+ \frac{g_0}{2} \hat{\psi}^\dagger(\r) \hat{\psi}^\dagger(\r) \hat{\psi}(\r) \hat{\psi}(\r)
\bigg\},    
\end{split}
\label{eq_H}
\end{equation}
where $\hat{\psi}^{(\dagger)}(\r)$ annihilates (creates) a boson of mass $m$ at position $\r=(x,y)$ in the 2D plane, and $\mu$ is the chemical potential. The coupling constant $g_0$ describes the effective contact interaction between bosons in the plane. For cold atoms confined along the third direction by a harmonic potential with typical energy $\omega_z=1/ml_z^2$, it is related to the ratio between the 3D s-wave scattering length $a_3$ and $l_z$, with the dimensionless coupling constant $m g_0\propto a_3/l_z$ \cite{Petrov00a,Lim08}. 

The above Hamiltonian correctly captures the physics at low enough temperature $T\ll \omega_z$ and density $n$ (corresponding to $\mu\ll \omega_z$). Furthermore, the contact interaction being ill-defined in 2D, we regularize the theory by introducing an ultraviolet cut-off $\Lambda\propto l_z^{-1}$. The model is thus valid only at small momenta $|\q|\ll \Lambda$ and large distances $|\r|\gg l_z$. At higher energies/momenta, a proper description of the confined 3D gas is needed.

That $m g_0$ is dimensionless implies that the classical theory of the 2D Bose gas is scale invariant under rescaling of space $\r\to s \r$ and energies $E\to E/s^2$, with the bosonic field having the scaling of the inverse of a length. At the quantum level, the regularization needed to make sense of the contact interaction breaks this scale invariance. This (approximate) scale invariance can be understood from a RG point of view as follows, see \cite{ranconUniversalThermodynamicsTwodimensional2012} for a detailed discussion. The scale invariance of the 2D dilute Bose gas is induced by the presence of a quantum critical point at $\mu=T=0$, separating two phases, the vacuum ($\mu<0$) and the superfluid phase ($\mu>0$) \cite{Sachdev94a}. The phase transition is second order, and the whole phenomenology of continuous phase transitions comes into play. From an RG point of view, the critical point is Gaussian, with critical exponent $\nu_0=1/2$, anomalous dimension $\eta_0=0$, and dynamical exponent $z_0=2$. Dimension two is the upper critical dimension of the QPT between the superfluid and the vacuum, implying that the renormalized interaction $\tilde g(s)$, depending on the rescaling factor $s$, vanishes logarithmically \footnote{There is a factor of two difference with the definition of Ref.~\cite{ranconUniversalThermodynamicsTwodimensional2012}. The one adopted here is such that $\tilde g(s)\simeq mg_0$ at weak coupling.}
\begin{equation}
\tilde{g}(s) = -\frac{2\pi}{\ln\!\left(\dfrac{\Lambda a_2}{2s}\right) + C},
\end{equation}
with $C$ the Euler's constant, and $a_2$ the 2D s-wave scattering length
\begin{equation}
a_2 = \frac{2}{\Lambda} \exp\!\left(-\frac{2\pi}{m g_0} - C\right).
\end{equation}
This quantum critical point governs the physics at finite temperature and chemical potential, and implies scaling for, in particular, thermodynamic quantities. For instance, the pressure obeys the scaling form
\begin{equation}
    P(\mu,T,mg_0) = s^{-4} P(\mu s^{2},T s^{2},\tilde g(s)).
\end{equation}
Choosing $s^{-2}\propto T$, one gets the scaling law for the pressure
\begin{equation}
    P(\mu,T,mg_0) = T\lambda_{\rm dB}^{-2} \mathcal{F}(\mu/T,\tilde g(T)),
\end{equation}
with $\lambda_{\rm dB}=\sqrt{2\pi/mT}$ the thermal de Broglie wavelength, and the energy-dependent coupling constant
\begin{equation}
\tilde{g}(\epsilon) = -\frac{2\pi}{\ln\!\left(\frac12\sqrt{2m a_2^2 \epsilon}\right) + C}=\frac{m g_0}{1-\frac{m g_0}{4\pi}\ln(\frac{2m\epsilon}{\Lambda^2})}.
\label{eq:couplage_g_epsilon}
\end{equation}
Similarly, the density $n=\partial_\mu P$ and the superfluid density $n_s$  take the universal form
\begin{align}
n(\mu,T,g_0) &= \lambda_{\rm dB}^{-2}
\mathcal D\!\left(\frac{\mu}{T},\tilde g(T)\right),\\
    n_s(\mu,T,g_0) &= \lambda_{\rm dB}^{-2}
\mathcal D_s\!\left(\frac{\mu}{T},\tilde g(T)\right),
\end{align}
with their corresponding universal functions $\mathcal D$ and $\mathcal D_s$.

These scaling relations hold for $|\mu|,T\ll\Lambda^2/2m$. They establish the universality of the dilute Bose gas thermodynamics near the quantum critical point: in two dimensions, the thermodynamics is determined by the mass $m$, the scattering length $a_2$, and the universal functions $\mathcal F$, $\mathcal D$, and $\mathcal D_s$, independently of the microscopic details. These results also apply to other dilute two-dimensional bosonic systems, such as dilute magnons~\cite{giamarchi_2008,Rancon14a} or cold atoms in optical lattices~\cite{fisher_1989,rancon_2011,ranconUniversalThermodynamicsTwodimensional2012}.

We note that the logarithmic dependence on the energy of $\tilde g(\epsilon)$ is important and needed to recover the correct equation of state of the ground state of the 2D Bose gas, which involves logarithms \cite{Mora09}.
In the weakly interacting limit, $mg_0\ll 1$, the renormalized interaction $\tilde g(\epsilon)$ barely varies with the energy, except at exponentially small energies (i.e. for $\epsilon\ll \frac{\Lambda^2}{2m}e^{-\frac{4\pi}{m g_0}}$). The phase-space pressure $\mathcal P=\lambda_{\rm dB}^2 P/T$ becomes approximately scale invariant and is parametrized by one number, the dimensionless interaction $m g_0$
\begin{equation}
    \mathcal P (\mu,T)\simeq\mathcal F(\mu/T,m g_0),
\end{equation}
the dependence on which we leave implicit from now on.

Computing the scaling function $\mathcal F$ explicitly is a difficult nonperturbative problem owing to the strong phase fluctuations of the 2D Bose gas at finite temperature. In the next section, we show how the FRG provides a precise calculation at both weak and strong coupling, throughout the phase diagram and across the superfluid transition.

\section{FRG implementation}
\label{sec_FRG}
\label{sec:implementationFRG}

The FRG approach has been extensively applied to bosonic quantum systems
\cite{dupuis_2007,wetterich_2008,floerchinger_2008,floerchinger_2009,dupuis_2009a,dupuis_2009b,sinner_2009,sinner_2010,rancon_2011,Rancon11a,ranconUniversalThermodynamicsTwodimensional2012,Rancon14a,Krieg17,isaule_2018,isaule_2020,Rancon23}. We focus here on the aspects specific to the present work.

\subsection{FRG for the dilute Bose gas}

The starting point is the action associated with Eq.~\eqref{eq_H},
\begin{equation}
S[\psi^*,\psi] = \int_0^\beta d\tau \int d^2r \left\{\psi^*\partial_\tau \psi
+ \psi^*\left( -\frac{\nabla^2}{2m} - \mu \right) \psi
+ \frac{g_0}{2} |\psi|^4\right\}.
\label{eq_action_boson}
\end{equation}
where $\beta=1/T$. Decomposing the complex field $\psi(\bm r,\tau)$ into two real components,
\begin{equation}
\psi(\bm r,\tau) = \frac{1}{\sqrt{2}}[\psi_1(\bm r,\tau) +
i\psi_2(\bm r,\tau)],
\label{eq_realfield_boson}
\end{equation}
with periodic boundary conditions $\psi(\bm r,\tau)=\psi(\bm r,\tau+\beta)$, gives
\begin{equation}
S[\bm{\psi}]
=
\int_0^\beta d\tau \int d^2r\,
\left[
\frac{i}{2}\,
\epsilon_{ij}\psi_i\partial_\tau\psi_j
+\frac{1}{4m}\,
(\nabla\bm{\psi})^2
-\frac{\mu}{2}\,
\bm{\psi}^2
+\frac{g_0}{8}\,
(\bm{\psi}^2)^2
\right],
\label{eq:action_ON_boson}
\end{equation}
where $\bm\psi=(\psi_1,\psi_2)$ and $\epsilon_{ij}$ is the antisymmetric Levi-Civita tensor.

The FRG constructs a family of models indexed by a momentum scale $k$, interpolating between a microscopic scale $\Lambda$ and zero~\cite{berges_2002,delamotte_2012,dupuis_2021}. This is achieved by adding a regulator term to the microscopic action,
\begin{equation}
\Delta S_k = \int_0^\beta d\tau \int d^2 r\int d^2 r'\,
\psi^*(\bm r,\tau)\, R_k(\bm r-\bm r')\, \psi(\bm r',\tau).
\label{eq:terme_regulateur}
\end{equation}
In momentum space, the cutoff function $R_k(\bm q)$ vanishes at $k=0$, so that $S+\Delta S_{k=0}$ recovers the original microscopic model. At finite $k$, $R_k(\bm q)$ suppresses low-momentum fluctuations with $|\bm q|\lesssim k$ and acts as an infrared regulator. At the microscopic scale $\Lambda$, it is chosen so that the model with action $S+\Delta S_\Lambda$ is exactly solvable: for $R_\Lambda(\bm q)\to\infty$ all fluctuations are frozen, and the mean-field approximation becomes exact. The regulator is independent of imaginary time, as justified below, and takes the form
\begin{equation}
R_k(\bm p) = Z_{x,k}\bm p^2\, r\!\left(\frac{\bm p^2}{k^2}\right),
\end{equation}
where $Z_{x,k}$ is defined below. The central quantity is the scale-dependent effective action $\Gamma_k$, defined by a modified Legendre transform of $\ln \mathcal{Z}_k[J^*,J]$,
\begin{equation}
\Gamma_k[\phi^*,\phi] = -\ln \mathcal{Z}_k[J^*,J]
+ \int_0^\beta d\tau \int d^2 r\, (J^*\phi+\mathrm{c.c.})
- \Delta S_k[\phi^*,\phi],
\label{eq:action_effective_Gammak}
\end{equation}
where $J(\bm r,\tau)$ is a complex source and $\mathcal{Z}_k$ is the generating functional in the presence of the regulator,
\begin{equation}
\mathcal{Z}_k[J^*,J]=\int\mathcal D\psi\mathcal D\psi^*\,
 e^{-S[\psi^*,\psi]-\Delta S_k[\psi^*,\psi]
 +\int_0^\beta d\tau\int d^2r\,(J^*\psi+\psi^*J)},
\end{equation}
while
\begin{equation}
\phi(\bm r,\tau) = \frac{\delta \ln \mathcal{Z}_k[J^*,J]}{\delta J^*(\bm r,\tau)}
= \langle \psi(\bm r,\tau)\rangle
\label{eq:parametre_ordre}
\end{equation}
is the superfluid order parameter. It is convenient to introduce the real and imaginary parts of the order parameter $\phi(\bm r,\tau) = \frac{1}{\sqrt{2}}[\phi_1(\bm r,\tau) +
i\phi_2(\bm r,\tau)]$.

At the microscopic scale $k=\Lambda$, the model is described by mean-field theory, and the effective action coincides with the microscopic action, $\Gamma_\Lambda[\phi^*,\phi]=S[\phi^*,\phi]$, which depends on $m$, $\mu$, $g_0$. The effective action at any scale $k$ is obtained by integrating the exact flow equation~\cite{wetterich_1993,ellwanger_1994,morris_1994}
\begin{equation}
\partial_k \Gamma_k[\phi^*,\phi] = \frac{1}{2}\mathrm{Tr}
\left\{\partial_k R_k
\left(\Gamma_k^{(2)}[\phi^*,\phi]+R_k\right)^{-1}\right\},
\label{eq:flot_wetterich_boson}
\end{equation}
where $\Gamma_k^{(2)}$ is the second functional derivative of $\Gamma_k$. In Fourier space, the trace in Eq.~\eqref{eq:flot_wetterich_boson} includes momenta, Matsubara frequencies, and the two components of the complex field $\phi$.

At the end of the flow, $\Gamma_{k=0}[\phi^*,\phi]=\Gamma[\phi^*,\phi]$ is the full effective action, from which the grand potential and all thermodynamic properties can be obtained. Temperature enters through the Matsubara frequencies in the flow equation. The effective action also inherits the dependence on microscopic parameters through the initial condition.

\subsection{Properties of the effective action}

The main quantities extracted from the effective action are the effective potential $U_k(\phi^*,\phi)$ and the two-point vertex $\Gamma_k^{(2)}[\phi^*,\phi]$. Their forms are strongly constrained by the symmetries of the microscopic action~\eqref{eq_action_boson}, provided that the regulator also preserves these symmetries. These constraints also suggest a physically motivated ansatz for solving the flow equation~\eqref{eq:flot_wetterich_boson}~\cite{dupuis_2009a, dupuis_2009b,dupuis_2021}.

The action is invariant under global $U(1)$ transformations
\begin{equation}
\psi^{(*)}(\bm r,\tau)\to e^{(\mp)i\alpha}\psi^{(*)}(\bm r,\tau),
\label{eq:u1}
\end{equation}
so that the effective potential is also $U(1)$ invariant and depends only on $\rho=(\phi_1^2+\phi_2^2)/2$. 
The equilibrium state is obtained by minimizing the effective potential, $U_k'(\rho_{0,k})=0$ if $\rho_{0,k}>0$, $\rho_{0,k}=0$ otherwise. It is characterized by the running condensate density $\rho_{0,k}$ and thermodynamic potential $U_k(\rho_{0,k})$. They give the condensate density and the Gibbs free energy in the limit $k=0$.

Moreover, $U(1)$ invariance implies that the two-point vertex in a constant field has the general form~\cite{dupuis_2009a}
\begin{equation}
\Gamma_{k,ij}^{(2)}(q;\phi) = \delta_{ij}\Gamma_{k,A}(q;\rho)
+ \phi_i\phi_j\,\Gamma_{k,B}(q;\rho)
+ \epsilon_{ij}\,\Gamma_{k,C}(q;\rho),
\label{eq:Gamma2_u1}
\end{equation}
where $q=(\bm q,i\omega_n)$, $\phi=(\phi_1,\phi_2)^T$ is the two-component real field, and $\delta_{ij}$ and $\epsilon_{ij}$ are the Kronecker and antisymmetric tensors, respectively. The functions $\Gamma_{k,A}$, $\Gamma_{k,B}$, and $\Gamma_{k,C}$ depend on the field only through $\rho$. The effective potential and the two-point vertex are related by
\begin{equation}
\Gamma_{k,ij}^{(2)}(q=0;\phi) = \frac{\partial^2 U_k(\rho)}{\partial\phi_i\partial\phi_j}
= \delta_{ij}U_k'(\rho) + \phi_i\phi_j\,U_k''(\rho).
\end{equation}

In addition to $U(1)$ invariance, the microscopic action~\eqref{eq_action_boson} has the discrete symmetries of time reversal $\psi(\bm r,\tau) \to \psi^*(\bm r,-\tau)$ and spatial inversion $\psi(\bm r,\tau) \to \psi(-\bm r,\tau)$,
which imply
\begin{equation}
\begin{aligned}
\Gamma_{k,ij}^{(2)}(\bm q,i\omega_n;\phi) &=
\Gamma_{k,ij}^{(2)}(-\bm q,i\omega_n;\phi), \\
\Gamma_{k,ij}^{(2)}(\bm q,i\omega_n;\phi) &=
(2\delta_{ij}-1)\,\Gamma_{k,ij}^{(2)}(\bm q,-i\omega_n;\phi^*),
\end{aligned}
\label{eq:Gamma2_parite}
\end{equation}
where $\phi^*=(\phi_1,-\phi_2)^T$. Equations~\eqref{eq:Gamma2_u1} and~\eqref{eq:Gamma2_parite} then give
\begin{equation}
\begin{aligned}
\Gamma_{k,A}(q;\rho) &= \Gamma_{k,A}(-q;\rho)
= \Gamma_{k,A}(\bm q,-i\omega_n;\rho), \\
\Gamma_{k,B}(q;\rho) &= \Gamma_{k,B}(-q;\rho)
= \Gamma_{k,B}(\bm q,-i\omega_n;\rho), \\
\Gamma_{k,C}(q;\rho) &= -\Gamma_{k,C}(-q;\rho)
= -\Gamma_{k,C}(\bm q,-i\omega_n;\rho).
\end{aligned}
\end{equation}

Inverting the two-point vertex gives the scale-dependent propagator,
\begin{equation}
G_k[\phi^*, \phi] = -\left(\Gamma_k^{(2)}[\phi^*, \phi] +
R_k\right)^{-1},
\label{eq:Gk_definition}
\end{equation}
whose tensor structure can be decomposed into longitudinal and transverse components, $G_{LL,k}$,  $G_{TT,k}$ and $G_{LT,k}$. Symmetry requires $G_{ij,k}$ to have the same form as Eq.~\eqref{eq:Gamma2_u1},
\begin{equation}
G_{ij,k}(q;\phi) = \frac{\phi_i\phi_j}{2\rho}G_{LL,k}(q;\rho)
+ \left(\delta_{ij}-\frac{\phi_i\phi_j}{2\rho}\right)G_{TT,k}(q;\rho)
+ \epsilon_{ij}G_{LT,k}(q;\rho),
\end{equation}
with
\begin{equation}
\begin{aligned}
G_{LL,k}(q;\rho) &= -\frac{\Gamma_{A,k}(q;\rho)+ R_k(\mathbf q)}{D_k(q;\rho)}, \\
G_{TT,k}(q;\rho) &= -\frac{\Gamma_{A,k}(q;\rho)+ R_k(\mathbf q)+2\rho\,\Gamma_{B,k}(q;\rho)}{D_k(q;\rho)}, \\
G_{LT,k}(q;\rho) &= \frac{\Gamma_{C,k}(q;\rho)}{D_k(q;\rho)},
\end{aligned}
\end{equation}
where $D_k=\det\left(\Gamma_k^{(2)}+ R_k\right)$.
These expressions recover the usual structure of the bosonic propagator in terms of normal and anomalous correlators \cite{dupuis_2009a, dupuis_2009b}. Furthermore, the gaplessness of the propagator at $k=0$, a consequence of the  Hugenholtz--Pines theorem~\cite{hugenholtz_1959}, is automatically enforced: $D_{k=0}(q=0;\rho_{0,k=0})\propto U'_{k=0}(\rho_{0,k=0})=0$ in the ordered phase.

\subsection{LPA\texorpdfstring{$'\times$DE$_2$}{'xDE2} approximation}
\label{subsec:LPApDE2}

We go beyond previous studies by extending the derivative expansion and avoiding a field expansion. At finite temperature, thermodynamics is dominated by long-wavelength classical fluctuations, whereas quantum fluctuations associated with nonzero Matsubara frequencies ($\omega_n=2\pi nT$, $n\neq0$) acquire a finite mass and renormalize the effective potential and kinetic coefficients at larger RG scales. Integrating out these quantum modes generates an effective classical $O(2)$ model governing the long-wavelength physics at $k\ll\lambda_{\rm dB}^{-1}$~\cite{Prokofev01,floerchinger_2009, popov_1983}. A regulator independent of imaginary time is therefore sufficient at nonzero temperature and allows for the Matsubara sums to be performed analytically. The infrared regulator ensures that the vertices are analytic for $|\bm q|\ll k$, which justifies a second-order derivative expansion for the classical sector. The quantum sector can be treated within the so-called LPA$'$, since we focus on the thermodynamics, which is dominated by classical spatial fluctuations. We denote this hybrid scheme by LPA$'\times$DE$_2$, with the ansatz
\begin{equation}
\Gamma_{k}[\phi^*, \phi] = \int_0^\beta d\tau \int d^2 r \left\{
Z_{\tau1,k}\phi^{*} \partial_\tau \phi
+ Z_{x,k}(\rho) \left| \nabla \phi \right|^{2}+ \frac14Y_{x,k}(\rho) \left( \nabla \rho \right)^{2}
+ U_k(\rho) \right\}.
\label{eq:ansatz}
\end{equation}
Although the microscopic action is polynomial in $\rho$, the flow generates a generic field dependence in these three functions. In the classical regime, the canonical field dimension vanishes in $d=2$. Thus, there is a priori no justification for neglecting high-order terms in $\rho$~\cite{jakubczyk_2014}. We therefore retain the full field dependence of the classical-sector functions $U_k$, $Z_{x,k}$, and $Y_{x,k}$.

By contrast, the quantum sector is treated within LPA$'$, and $Z_{\tau1,k}$ is taken to be field independent and evaluated at the minimum $\rho_{0,k}$ of the effective potential, see Eq.~\eqref{eq:flot_Ztau1} below. 
Previous studies~\cite{wetterich_2008,dupuis_2009a,dupuis_2009b,ranconUniversalThermodynamicsTwodimensional2012} included a second-order time-derivative term in the ansatz~\eqref{eq:ansatz},
\begin{equation}
\int_0^\beta d\tau \int d^2 r\, Z_{\tau,k}\,|\partial_\tau\phi|^2,
\label{eq:terme_Ztau}
\end{equation}
while also expanding all other functions around $\rho_{0,k}$. This term, while absent at $k=\Lambda$, is particularly important to describe the non-perturbative low-energy properties below a Ginzburg scale $\omega_G\sim (mg_0)^2 g_0n$ \cite{dupuis_2009a,dupuis_2009b}. At weak coupling, this scale is much smaller than the typical interaction energy $g_0n\sim \mu$, of the same order as the temperatures considered here. Furthermore, since only the quantum modes involve $Z_{\tau,k}$, and since those decouple from the flow equations for $T\gg k^2/2m\gg \omega_G$, we can safely dismiss $Z_{\tau,k}$ when computing the thermodynamics. In Sec.~\ref{sec:convergence_lpade2}, we check that this omission does not affect the thermodynamic properties appreciably.

In the ordered phase, for a uniform field $\phi_i=\delta_{i1}\sqrt{2\rho}$, the two-point vertex reads
\begin{equation}
\Gamma_{k}^{(2)} =
\begin{pmatrix}
\Gamma_{k,A}+2\rho\,\Gamma_{k,B} & \Gamma_{k,C} \\
-\Gamma_{k,C} & \Gamma_{k,A}
\end{pmatrix},
\end{equation}
with
\begin{equation}
\begin{aligned}
\Gamma_{k,A}(q;\rho) &= \bm q^2 Z_{x,k}(\rho)+ U_k'(\rho), \\
\Gamma_{k,B}(q;\rho) &= \frac{1}{2}\bm q^2 Y_{x,k}(\rho)+ U_k''(\rho), \\
\Gamma_{k,C}(q;\rho) &= Z_{\tau1,k}\omega_n,
\end{aligned}
\end{equation}
in agreement with the symmetries established above.

The running anomalous dimension is defined by
\begin{equation}
\eta_k = -k\partial_k\ln Z_{x,k},
\label{eq:dimetaboson}
\end{equation}
where, with a slight abuse of notation, $Z_{x,k}\equiv Z_{x,k}(\rho=0)$ denotes the field renormalization factor evaluated at the arbitrary renormalization point $\rho_r=0$. The physical anomalous dimension is $\eta=\lim_{k\to0}\eta_k$.

\subsection{Flow equations}

In terms of the propagator~\eqref{eq:Gk_definition}, the Wetterich equation~\eqref{eq:flot_wetterich_boson} reads ($t=\ln(k/\Lambda))$
\begin{equation}
\partial_t\Gamma_k[\bm\phi]=-\frac12\,\mathrm{Tr}\left\{\partial_tR_k\,
G_k[\bm\phi]\right\}.
\label{eq:flot_wetterich_G}
\end{equation}
The flow of the effective potential is obtained for a uniform and stationary field,
\begin{equation}
\partial_t U_k(\rho)=-\frac{1}{2\beta V}\,\mathrm{Tr}\left\{\partial_tR_k\,
G_k\right\}\Big|_{\bm\phi\ \rm cst},
\label{eq:flot_U_general}
\end{equation}
whose integration is discussed in Sec.~\ref{subsec:discretization}. The minimum of the effective potential also evolves during the flow. As long as $\rho_{0,k}>0$, the condition $U_k'(\rho_{0,k})=0$ for all $k$ gives
\begin{equation}
\partial_t\rho_{0,k}
=-\frac{\partial_tU_k'(\rho_{0,k})}
{U_k''(\rho_{0,k})}.
\label{eq:flot_rho0}
\end{equation}
If $\rho_{0,k}$ vanishes at a finite scale, the minimum remains at the origin thereafter, with $U_k'(0)>0$, and the system is in the normal phase. In practice, since the full functional dependence of $U_k$ is retained, $\rho_{0,k}$ is determined at each integration step as the zero of $U_k'(\rho)$.

Taking two functional derivatives of the Wetterich equation with respect to the field yields the flow of the two-point vertex in a uniform and stationary field~\cite{dupuis_2009a},
\begin{equation}
\begin{aligned}
\partial_t \Gamma_{k,ij}^{(2)}(p;\bm\phi)
={}&-\frac{1}{2}\sum_{q,i_1,i_2}
\tilde{\partial}_t G_{k,i_1i_2}(q;\bm\phi)\,
\Gamma_{k,iji_2i_1}^{(4)}(p,-p,q,-q;\bm\phi)
\\
&-\frac{1}{2}\sum_{q,i_1,\ldots,i_4}
\Big\{
\Gamma_{k,ii_2i_3}^{(3)}(p,q,-p-q;\bm\phi)\,
\Gamma_{k,ji_4i_1}^{(3)}(-p,p+q,-q;\bm\phi)
\\
&\qquad\qquad\times
G_{k,i_3i_4}(p+q;\bm\phi)\,\tilde{\partial}_t G_{k,i_1i_2}(q;\bm\phi)
+\big(p\leftrightarrow -p,\; i\leftrightarrow j\big)
\Big\},
\end{aligned}
\label{eq:flot_Gamma2}
\end{equation}
where $p=(\bm p,i\omega_n)$ and $q=(\bm q,i\omega_m)$ are the external and internal momentum-frequency variables, respectively, $\sum_q=\beta^{-1}\sum_{\omega_m}\int_{\bm q}$ and $\int_{\bm q}=\int \frac{d{\bm q}}{(2\pi)^d}$, and the operator
\begin{equation}
\tilde{\partial}_t
=(\partial_t R_k)\,\frac{\partial}{\partial R_k}
+(\partial_t R_k')\,\frac{\partial}{\partial R_k'},
\label{eq:dt_tilde}
\end{equation}
with $R_k'(\bm q)\equiv\partial_{\bm q^2}R_k(\bm q)$, acts only on the dependence of the propagators on $R_k$. The flow of $\Gamma_k^{(2)}$ involves $\Gamma_k^{(3)}$ and $\Gamma_k^{(4)}$,
and so on, such that the Wetterich equation is equivalent to an infinite hierarchy of coupled equations. The ansatz~\eqref{eq:ansatz} closes this hierarchy, with the vertices $\Gamma_k^{(3)}$ and $\Gamma_k^{(4)}$ obtained by taking functional derivatives of the ansatz with respect to the field. The flows of $Z_{x,k}(\rho)$, $Y_{x,k}(\rho)$, and $Z_{\tau1,k}$ are then obtained by projecting Eq.~\eqref{eq:flot_Gamma2} onto the ansatz,
\begin{align}
\partial_t Z_{x,k}(\rho) &= \lim_{\bm p\to0}
\partial_{\bm p^2}\,\partial_t\Gamma_{k,A}(\bm p,\omega_n=0;\rho),
\label{eq:flot_Zx} \\[1ex]
\partial_t Y_{x,k}(\rho) &= 2\lim_{\bm p\to0}
\partial_{\bm p^2}\,\partial_t\Gamma_{k,B}(\bm p,\omega_n=0;\rho),
\label{eq:flot_Yx} \\[1ex]
\partial_t Z_{\tau1,k} &= \lim_{\omega_n\to0}
\partial_{\omega_n}\,\partial_t\Gamma_{k,C}(\bm p=0,\omega_n;\rho_{0,k}).
\label{eq:flot_Ztau1}
\end{align}
The full flow equations are given in App.~\ref{sec_Appendix1}. The initial condition $\Gamma_{k=\Lambda}[\bm\phi]=S[\bm\phi]$ gives
\begin{equation}
\begin{aligned}
U_{\Lambda}(\rho) &= -\mu\rho + \frac{g_0}{2}\rho^2, \\
Z_{x,\Lambda}(\rho) &= Z_{\tau1,\Lambda} = 1, \\
Y_{x,\Lambda}(\rho) &= 0.
\end{aligned}
\label{eq:initial_conditions}
\end{equation}
We set $2m=1$ from now on, without loss of generality.

\subsection{Numerical resolution of the flow equations \label{subsec:discretization}}

The main difficulty in solving the flow equations comes from the flow of the effective potential. Since it is a true Legendre transform at $k=0$, it must be a convex function in that limit. This return to convexity is known to be subtle to resolve numerically \cite{Pelaez16}. Using ideas from fluid dynamics, this problem has been greatly simplified for the classical $O(N)$ model at LPA \cite{grossi_2023,ihssen_2023,koenigstein_2022,koenigsteinNumericalFluidDynamics2022a}. 
We therefore extend to the dilute Bose gas the upwind/downwind discretization introduced there. 

The equation for $V_k(\rho)=\partial_\rho U_k(\rho)$ takes the form of a conservation law,
\begin{equation}
\partial_t V_k(\rho) = \partial_\rho f_{k,LL}(\rho) + \partial_\rho f_{k,TT}(\rho),
\label{eq:flot_V_k_boson}
\end{equation}
with the longitudinal and transverse fluxes $f_{k,LL}$ and $f_{k,TT}$, respectively,
\begin{equation}
f_{k,LL/TT}(\rho) = -\frac{1}{2\beta}\sum_{\omega_n}
\int_{\bm p} \partial_t R_k(\bm p)\,G_{LL/TT,k}(p;\rho).
\label{eq:flux_LL_TT}
\end{equation}
At each step in $\rho$ and $t$, evaluating $f_{k,LL}$ and $f_{k,TT}$ requires the integration over the momentum $\bm p$ and the sum over the Matsubara frequencies $\omega_n$. The former is performed numerically by Gauss-Legendre quadrature, while the latter is computed analytically.

The structure of the propagators nevertheless differs from that of the $O(N)$ model. Denoting, as for the $O(N)$ model, the transverse and longitudinal vertices
\begin{equation}
\begin{aligned}
\Gamma^{(2)}_{k,T}(\bm p;\rho) &=  Z_{x,k}(\rho)\bm p^2+V_k(\rho),\\
\Gamma^{(2)}_{k,L}(\bm p;\rho) &= \big(Z_{x,k}(\rho)+\rho Y_{x,k}(\rho)\big)\bm p^2
+V_k(\rho)+2\rho\,\partial_\rho V_k(\rho),
\end{aligned}
\label{eq:GammaT_GammaL}
\end{equation}
the propagators read
\begin{align}
G_{LL,k}(\bm p,\omega_n;\rho) &= -\frac{\Gamma^{(2)}_{k,T}(\bm p;\rho)+R_k(\bm p)}{D_k(\bm p,\omega_n;\rho)},
\label{eq:G_LL_explicite} \\[1ex]
G_{TT,k}(\bm p,\omega_n;\rho) &= -\frac{\Gamma^{(2)}_{k,L}(\bm p;\rho)+R_k(\bm p)}{D_k(\bm p,\omega_n;\rho)},
\label{eq:G_TT_explicite} \\[1ex]
D_k(\bm p,\omega_n;\rho) &= \big(\Gamma^{(2)}_{k,T}+R_k\big)\big(\Gamma^{(2)}_{k,L}+R_k\big)
+Z_{\tau1,k}^2\,\omega_n^2 .
\label{eq:D_k_explicite}
\end{align}

In the classical $O(N)$ model, the longitudinal mass $V_k+2\rho\,\partial_\rho V_k$ appears only in the longitudinal propagator, where it plays the role of a diffusive term in the flow. For the Bose gas, the off-diagonal vertex $\Gamma_{k,C}=Z_{\tau1,k}\omega_n$ couples the longitudinal and transverse fluctuations of the nonzero Matsubara modes, so that $\Gamma^{(2)}_{k,L}$ enters both propagators. In $G_{LL,k}$, it appears only in the denominator $D_k$, and for $\omega_n=0$, $G_{LL,k}=-1/(\Gamma^{(2)}_{k,L}+R_k)$ recovers the longitudinal propagator of the $O(N)$ model. In $G_{TT,k}$, it appears in both the denominator and the numerator. These two dependences cancel exactly for $\omega_n=0$, where $G_{TT,k}=-1/(\Gamma^{(2)}_{k,T}+R_k)$ recovers the classical transverse propagator, so that the dependence of $G_{TT,k}$ on $\Gamma^{(2)}_{k,L}$ comes solely from the quantum sector.

Let us first recall how to deal with the return to convexity in the classical $O(N)$ model at LPA. There, $f_{k,TT}(V_k)$ accounts for the $N-1$ transverse Goldstone modes, with mass $V_k$ whose flow is purely convective since it depends only on $V_k$ itself and not on its derivatives. The term $f_{k,LL}(V_k+2\rho\partial_\rho V_k)$ accounts for the single longitudinal mode, with mass $V_k+2\rho\partial_\rho V_k$, and introduces the diffusive component.
The flow equation can be recast as a conservation law with advective and diffusive contributions 
\begin{equation}
    \partial_t V_k+\partial_\rho F=0,
    \label{eq:flux}
\end{equation}
with flux $F=-((N-1)f_{k,TT}+f_{k,LL})$.
For an advection equation of the form $\partial_t u+\partial_\rho F(u)=0$, the sign of $\partial_u F(u)$, which plays the role of a characteristic velocity, determines the direction of information propagation. A positive characteristic velocity transports information towards larger values of $\rho$, while a negative velocity transports it towards smaller values.
Since $f_{k,LL}$ and $f_{k,TT}$ are positive (note that $\partial_t R_k>0$) and are both decreasing functions of their argument, $F$ is an increasing function of $V_k$, corresponding to a positive characteristic velocity in $t$. Since the flow proceeds towards $t\to-\infty$, information therefore propagates, along the flow, from larger to smaller values of $\rho$ throughout the domain. A standard analysis shows that the derivative of the flux must be discretized using an upwind scheme that follows the flow of information. The derivative with respect to $\rho$ appearing in Eq.~\eqref{eq:flux} must therefore be discretized using larger values of $\rho$, through the forward operator $d_{\mathrm{u}}$.
The diffusive term $2\rho\partial_\rho V_k$ in the longitudinal mass requires a symmetric discretization, unlike advection, which favors a particular direction. One therefore combines the forward derivative used for the total flux with a downwind (backward) derivative $d_{\mathrm{d}}$ for this term. 

These discrete derivatives are defined as follows.
We discretize the field domain on an ordered grid of $N_\rho$ points,
\begin{equation}
\{\rho_i\}=\{\rho_1=0,\ \rho_2,\ldots,\rho_{N_\rho}=\rho_{\max}\}.
\end{equation}
We denote $V_{k,i}\equiv V_k(\rho_i)$. The forward and backward finite-difference operators are defined, for any discretized function $a_i$, by
\begin{equation}
\begin{split}
    (d_{\mathrm{u}}a)_i&=\frac{a_{i+1}-a_i}{\rho_{i+1}-\rho_i},\\
(d_{\mathrm{d}}a)_i&=\frac{a_i-a_{i-1}}{\rho_i-\rho_{i-1}}.
\end{split}
\end{equation}

We are now in a position to address the specifics of the Bose gas flow equations. As in the classical $O(N)$ model, the flux is positive and decreases for increasing $\rho$. Thus the discussion above applies. In the denominator $D_k$, the diffusive term is discretized with the downwind operator $d_d$. In the numerator of $G_{TT,k}$, by contrast, an increase in $\Gamma^{(2)}_{k,L}$ increases the flux $f_{k,TT}$. For this term, which has no classical counterpart, no analogous argument fixes a priori the direction of the discretization, and numerical tests show that an upwind scheme is required to ensure the stability of the integration down to the regime $k\to0$. Defining the two discretized longitudinal masses
\begin{align}
m^2_{-,i} &= V_{k,i}+2\rho_i(d_d V_k)_i, \\
m^2_{+,i} &= V_{k,i}+2\rho_i(d_uV_k)_i,
\label{eq:masses_discretisees_boson}
\end{align}
the discretized fluxes read
\begin{equation}
\begin{aligned}
f_{k,LL}\big(m^2_{-,i}\big) &= \frac{1}{2\beta}\sum_{\omega_n}\int_{\bm p}
\partial_tR_k(\bm p)\,
\frac{\Gamma^{(2)}_{k,T,i}+R_k}{D_{k,i}\big(m^2_{-,i}\big)},\\
f_{k,TT}\big(m^2_{-,i},m^2_{+,i}\big) &= \frac{1}{2\beta}\sum_{\omega_n}\int_{\bm p}
\partial_tR_k(\bm p)\,
\frac{\Gamma^{(2)}_{k,L,i}\big(m^2_{+,i}\big)+R_k}{D_{k,i}\big(m^2_{-,i}\big)},
\end{aligned}
\label{eq:flux_discretises_boson}
\end{equation}
with
\begin{equation}
\begin{aligned}
\Gamma^{(2)}_{k,T,i}&=\bm p^2Z_{x,k,i}+V_{k,i},\\
\Gamma^{(2)}_{k,L,i}\big(m^2\big)&=\bm p^2\big(Z_{x,k,i}+\rho_iY_{x,k,i}\big)+m^2,\\
D_{k,i}\big(m^2\big)&=\big(\Gamma^{(2)}_{k,T,i}+R_k\big)
\big(\Gamma^{(2)}_{k,L,i}(m^2)+R_k\big)
+Z_{\tau1,k}^2\,\omega_n^2 .
\end{aligned}
\end{equation}
The discretized flow then reads
\begin{equation}
\partial_t V_{k,i}
=\Big(d_\mathrm{u}\big[f_{k,LL}\big(m^2_{-}\big)
+f_{k,TT}\big(m^2_{-},m^2_{+}\big)\big]\Big)_i ,
\label{eq:flot_V_discretise_boson}
\end{equation}
with the boundary conditions $(\rho\, d_da)_1=0$ and $(d_u a)_{N_\rho}=(d_d a)_{N_\rho}$.

The integration of the flows of $Z_{x,k}$ and $Y_{x,k}$ does not encounter the stiffness associated with the return to convexity of the potential. A two-point centered scheme suffices, and we use $m_-^2$ in the denominator of propagators, and $m_+^2$ in the numerators. Note that $V_k''=\partial_\rho^3 U_k$ also appears in the flow of $Y_{x,k}$ and is likewise discretized with a centered scheme. The grid in $\rho$ is uniform and fixed along the flow, with $\rho_{\max}=2\rho_{0,\Lambda}$, where $\rho_{0,\Lambda}=\mu/g_0$ is the minimum of $U_\Lambda$ for positive chemical potential, Eq.~\eqref{eq:initial_conditions} (a lower bound on $\rho_{\max}$ is imposed for small or negative $\mu$). For instance, we typically use $N_\rho=65$ points for $mg_0=0.15$.

The flow equations are integrated with the Wetterich regulator,
\begin{equation}
R_k(\bm p) = Z_{x,k}\,\bm p^2\, r\!\left(\frac{\bm p^2}{k^2}\right),
\qquad
r(y) = \frac{\alpha}{e^{y}-1},
\label{eq:regulateur_exp}
\end{equation}
where $Z_{x,k}\equiv Z_{x,k}(\rho=0)$ is the field renormalization, Eq.~\eqref{eq:dimetaboson}, and $\alpha$ is set to $\alpha=2.1$, for which the flow is most stable numerically at long RG time. We have checked, by varying $\alpha$ around this value, that this choice does not affect the thermodynamics. The reliability of the thermodynamic results obtained with this choice is supported also a posteriori by the comparisons in Sec.~\ref{sec:convergence_lpade2}. The two regulators considered within LPA' yield similar equations of state, and extending the approximation to LPA$'\times$DE$_2$ produces only small corrections.

\begin{figure}[t]
\centering
\begin{minipage}[t]{0.495\textwidth}
    \centering
    \includegraphics[width=\linewidth]{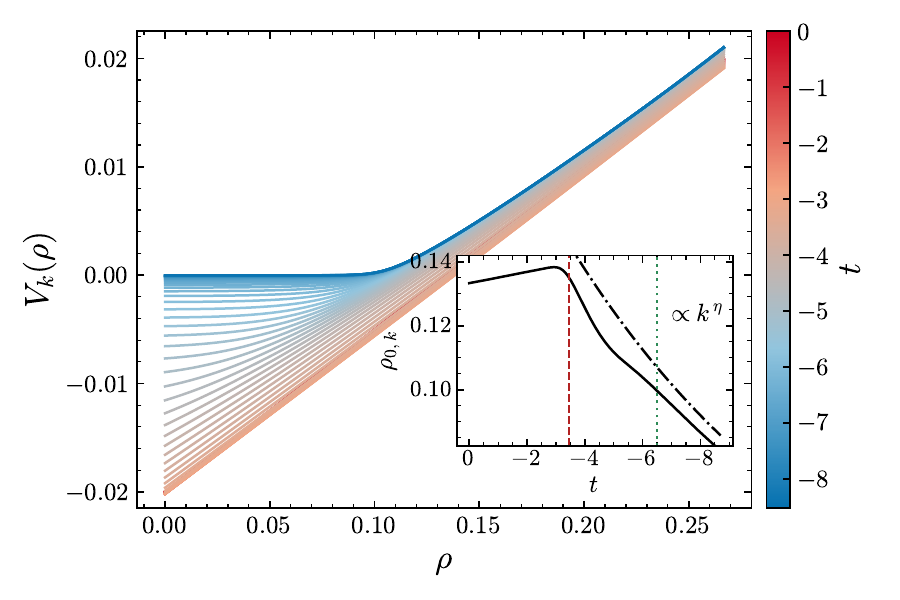}
\end{minipage}\hspace{0.01\textwidth}%
\begin{minipage}[t]{0.495\textwidth}
    \centering
    \includegraphics[width=\linewidth]{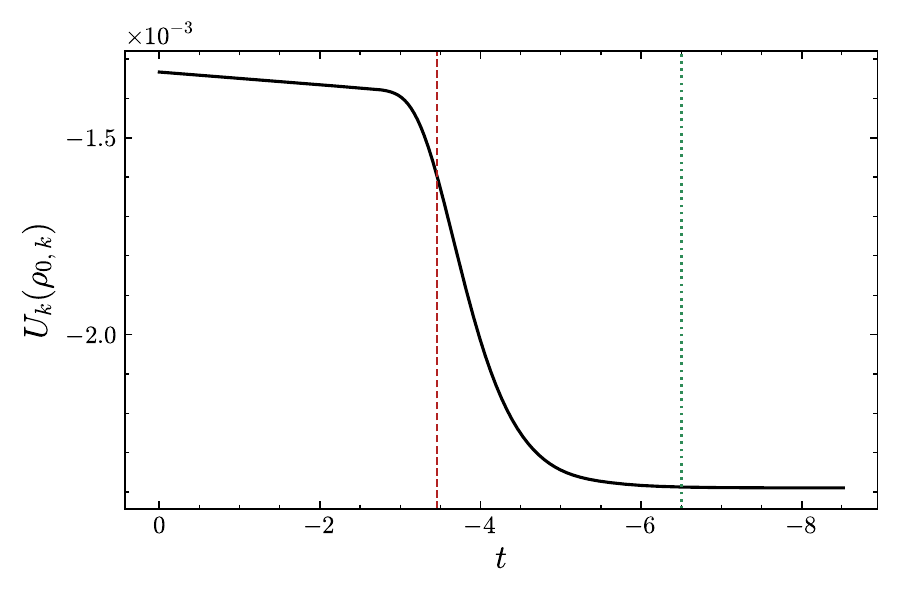}
\end{minipage}

\vspace{0.5em}

\begin{minipage}[t]{0.49\textwidth}
    \centering
    \includegraphics[width=\linewidth]{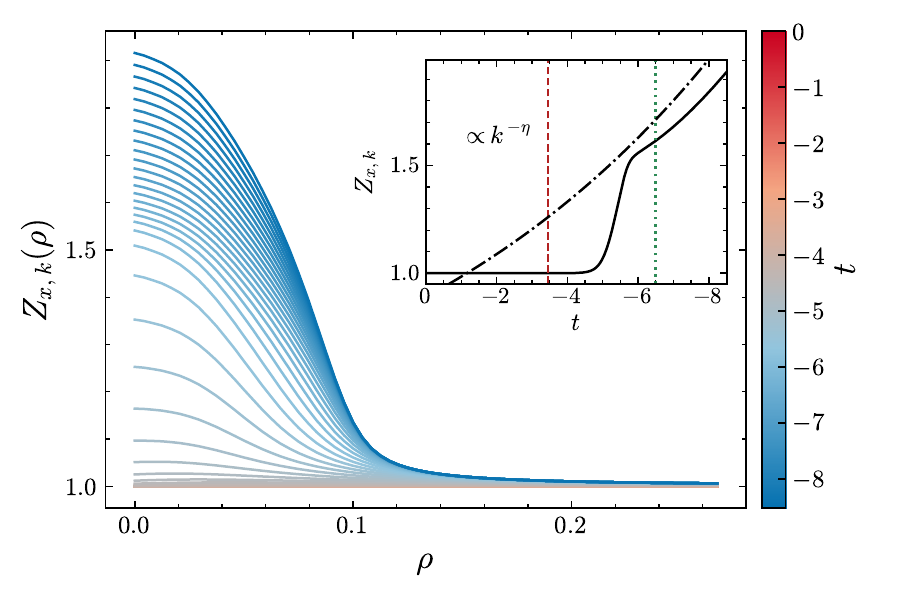}
\end{minipage}\hfill
\begin{minipage}[t]{0.49\textwidth}
    \centering
    \includegraphics[width=\linewidth]{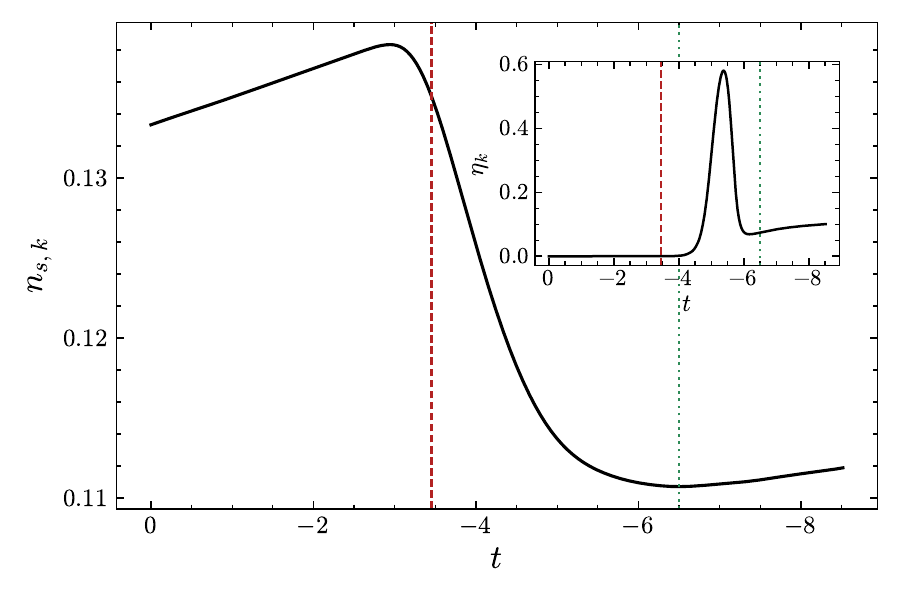}
\end{minipage}
\caption{Numerical integration of the flow for the dilute two-dimensional Bose gas at weak coupling, $mg_0=0.15$, in the low-temperature phase, with $\mu/T=0.2$. Top left: $V_k(\rho)$ develops a plateau for $\rho<\rho_{0,k}$ as $|t|$ increases, indicating the restoration of convexity. Colors indicate the value of $t$. The inset shows the running minimum $\rho_{0,k}$, which asymptotically follows the power law $\propto k^\eta$. Top right: $U_k(\rho_{0,k})$ reaches its physical value at the scale $k_{\rm conv}$, marked by the green dotted line. Bottom left: $Z_{x,k}(\rho)$ as a function of $\rho$, with the same color scale for $t$. The inset shows the field renormalization factor $Z_{x,k}$ as a function of $t$. Bottom right: running superfluid density $n_{s,k}$ and, in the inset, running anomalous dimension $\eta_{x,k}$, both as functions of $t$. In all four panels, the red dashed line marks the scale $k_{\rm dB}\propto\sqrt T$, separating the quantum regime from the effective classical regime.}
\label{fig:flot_boson_UZeta}
\end{figure}

Figure~\ref{fig:flot_boson_UZeta} shows the numerical integration for a given set of microscopic parameters $(\mu,T,mg_0,\Lambda)$. Unless stated otherwise, we choose $T/\Lambda^2/2m=10^{-3}$ fixed, and vary $\mu$. The flow of $V_k$ restores convexity as $k\to0$. The running minimum is determined by $V_k(\rho_{0,k})=0$, from which we obtain the running pressure $-U_k(\rho_{0,k})$. The latter already converges at a finite scale $k_{\rm conv}$, below which $U_k(\rho_{0,k})$ no longer evolves, which allows for the pressure to be extracted unambiguously, $P=-U_{k\lesssim k_{\rm conv}}(\rho_{0,k\lesssim k_{\rm conv}})$, without having to continue the flow down to $k=0$. A second characteristic scale, the inverse thermal de Broglie wavelength $k_{\rm dB}\propto\sqrt T$, marks the crossover from the quantum regime to the classical regime that dominates the thermodynamics, in which the nonzero Matsubara frequency modes have been integrated out~\cite{blaizot_2012}. Below $k_{\rm dB}$, the flow of $Z_{\tau1,k}$ decouples from that of the other coefficients, and the system behaves as an effective classical $O(2)$ model~\cite{floerchinger_2009}.

\subsection{Extracting the universal scaling functions from FRG}

The universal scaling functions can be obtained from the solution of the FRG flow equations, with initial conditions Eq.~\eqref{eq:initial_conditions} and a choice of temperature and chemical potential. The ultraviolet cut-off is chosen such that $T,\mu\ll \Lambda^2$ (see also below).

The pressure follows directly from the flow of $U_k$ through $P=-\lim_{k\to0}U_k(\rho_{0,k})$.
The phase-space pressure and density are then obtained from
\begin{align}
\mathcal{P}\!\left(\frac{\mu}{T},\tilde{g}(T)\right) &=
-\frac{\lambda^2_{\rm dB}}{T} \lim_{k\to 0} U_k(\rho_{0,k}),
\label{eq:pression_adim} \\
\mathcal{D}\!\left(\frac{\mu}{T},\tilde{g}(T)\right) &=
-\lambda^2_{\rm dB} \partial_\mu\lim_{k\to 0} 
U_k(\rho_{0,k}).
\label{eq:densite_adim}
\end{align}
Successive derivatives of the pressure with respect to $\mu$ and $T$ give all thermodynamic quantities. 

The superfluid density measures the stiffness of the system against a phase twist~\cite{fisher_1973}. It is obtained from the coefficient of the gradient term $n_{s,k}(\nabla\theta)^2$ in the low-momentum effective action for the phase $\theta$ of the superfluid order parameter. It can be read directly from the transverse component of the two-point vertex~\cite{jakubczyk_2014, dupuis_2009a, chaikin_1995},
\begin{equation}
\Gamma_{A,k}(\bm q,\omega_n=0;\rho_{0,k})
= \frac{n_{s,k}}{\rho_{0,k}}\bm q^2
\qquad (\bm q\to0).
\label{eq:nsdef}
\end{equation}
The physical superfluid density is therefore $n_s=\lim_{k\to0}n_{s,k}$, from which we obtain the superfluid phase-space density, 
\begin{align}
\mathcal{D}_s\!\left(\frac{\mu}{T},\tilde{g}(T)\right) &=
\lambda^2_{\rm dB}\lim_{k\to0}Z_{x,k}(\rho_{0,k})\, \rho_{0,k}.
\end{align}
The anomalous dimension is obtained from the flow of $Z_{x,k}(\rho)$ through the field renormalization factor $Z_{x,k}=Z_{x,k}(0)$. Equation~\eqref{eq:dimetaboson} implies $Z_{x,k}\propto k^{-\eta}$. The flow also displays the signatures of the BKT mechanism in the low-temperature phase~\cite{grater_1995,gersdorff_2001,jakubczyk_2014,ranconUniversalThermodynamicsTwodimensional2012}, see Sec.~\ref{sec:FRGandBKT} for a detailed discussion. The absence of true long-range order and spontaneous symmetry breaking implies $\lim_{k\to0}\rho_{0,k}=0$, whereas the running superfluid density $n_{s,k}$ and anomalous dimension $\eta_k$ remain finite at finite $k$.

\section{Robustness of the thermodynamics with respect to the approximation scheme}
\label{sec:convergence_lpade2}

The convergence of the derivative and field expansions has been studied in detail for $O(N)$ models, notably in Refs.~\cite{canet_2003,balog_2019,depolsi_2020}. Here, we assess the sensitivity of the thermodynamics to the approximation scheme by comparing the results obtained within the LPA$'\times$DE$_2$ and the LPA$'$.  
The latter was used in~\cite{ranconUniversalThermodynamicsTwodimensional2012} to compute the universal thermodynamics of the 2D Bose gas, with an additional  field expansion of the effective potential,
\begin{equation}
U_k(\rho) = U_{0,k} + \delta_k(\rho-\rho_{0,k}) +
\frac{\lambda_k}{2}(\rho-\rho_{0,k})^2,
\end{equation}
where $\delta_k=0$ when $\rho_{0,k}>0$, and a $\Theta$-regulator $R_k(\mathbf p)=Z_{x,k}(k^2-\mathbf p^2)\Theta(k^2-\mathbf p^2)$, $\Theta$ the Heaviside function. This LPA$'$ approximation sets $Y_{x,k}=0$ and treats $Z_{x,k}$ as field independent, evaluated at $\rho_{0,k}$. 

\begin{figure}[t]
\centering
\begin{minipage}[t]{0.495\textwidth}
    \centering
    \includegraphics[width=\linewidth]{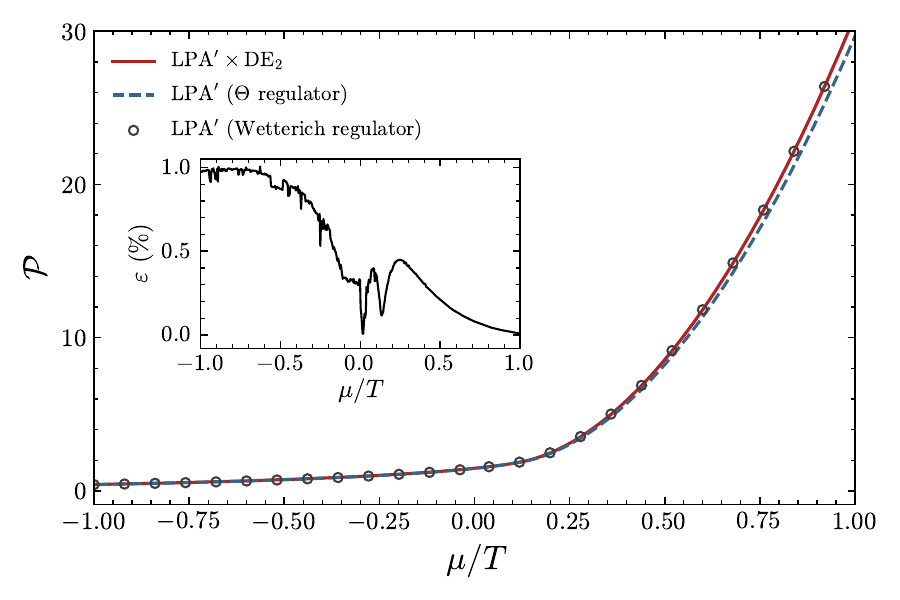}
\end{minipage}\hspace{0.01\textwidth}%
\begin{minipage}[t]{0.495\textwidth}
    \centering
    \includegraphics[width=\linewidth]{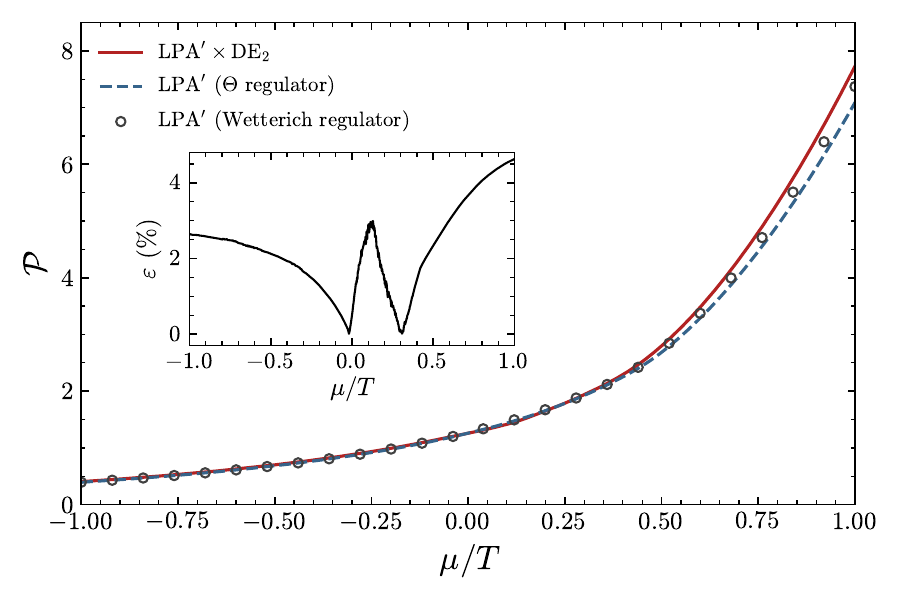}
\end{minipage}
\caption{Phase-space pressure $\mathcal{P}$ as a function of $\mu/T$ for $mg_0=0.11$ (left) and $mg_0=0.64$ (right), comparing LPA$'\times$DE$_2$ (solid red line), LPA$'$ with the $\Theta$ regulator (dashed blue line) and LPA$'$ with Wetterich's exponential regulator (open circles). The insets show the relative difference $\varepsilon$ between LPA$'\times$DE$_2$ and LPA$'$ with Wetterich's regulator.}
\label{fig:convergence_pression}
\end{figure}

Figure~\ref{fig:convergence_pression} shows this comparison for the pressure $P(\mu/T,mg_0)$ at weak coupling ($mg_0=0.11$, left panel) and strong coupling ($mg_0=0.64$, right panel). We also show LPA$'$ results using the Wetterich regulator Eq.~\eqref{eq:regulateur_exp} with $\alpha=2.1$. In both regimes, the agreement is good over the entire range of $\mu/T$ considered. The insets show that the relative difference $\varepsilon$ is always less than $5\,\%$, including near the BKT transition ($\mu/T\gtrsim0$). The close agreement between the two regulators within LPA', together with the small corrections introduced by LPA$'\times$DE$_2$, supports the robustness of the pressure with respect to the regulator and approximation choices considered here.

\begin{figure}[]
\centering
\includegraphics[width=0.63\textwidth]{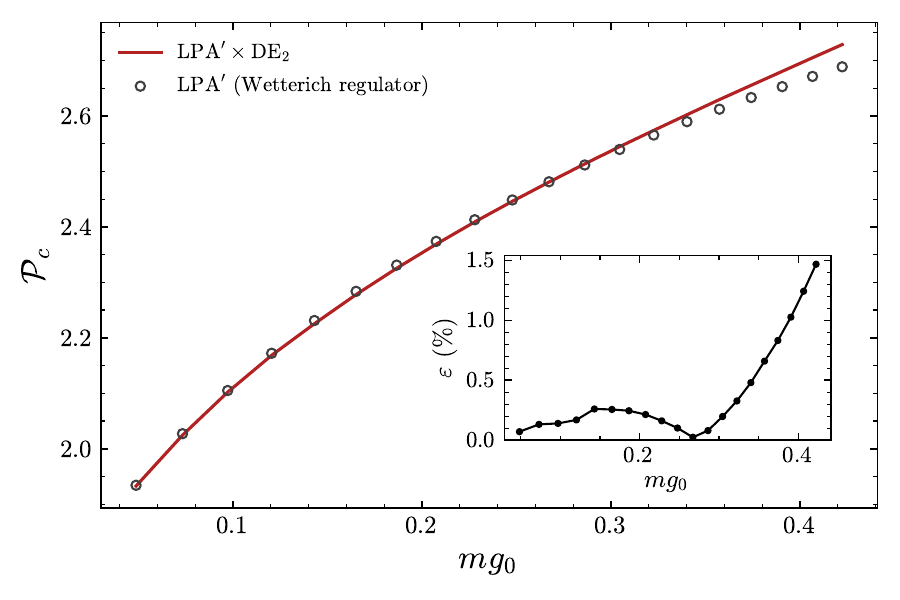}
\caption{Phase-space critical pressure $\mathcal{P}_c=\mathcal{P}(\mu_c/T)$ at the BKT transition as a function of $mg_0$, comparing LPA$'\times$DE$_2$ (solid red line) and LPA$'$ with Wetterich's exponential regulator (open circles). The inset shows the relative difference $\varepsilon$ between the two approximations.}

\label{fig:convergence_nc}
\end{figure}

Since the two approximations agree closely throughout the range of $\mu/T$, we extend the comparison to a particular point in the phase diagram: the BKT critical point $\mu_c/T$. We compute the LPA$'$ and LPA$'\times$DE$_2$ equations of state for different interaction strengths; the extraction of $\mu_c$ is discussed in Sec.~\ref{sec:FRGandBKT}. Figure~\ref{fig:convergence_nc} shows the critical pressure $\mathcal P_c$ as a function of $m g_0$. The agreement remains excellent across the entire interaction range, with corrections at the percent level even at strong coupling.

\begin{figure}[]
\centering
\includegraphics[width=0.6\textwidth]{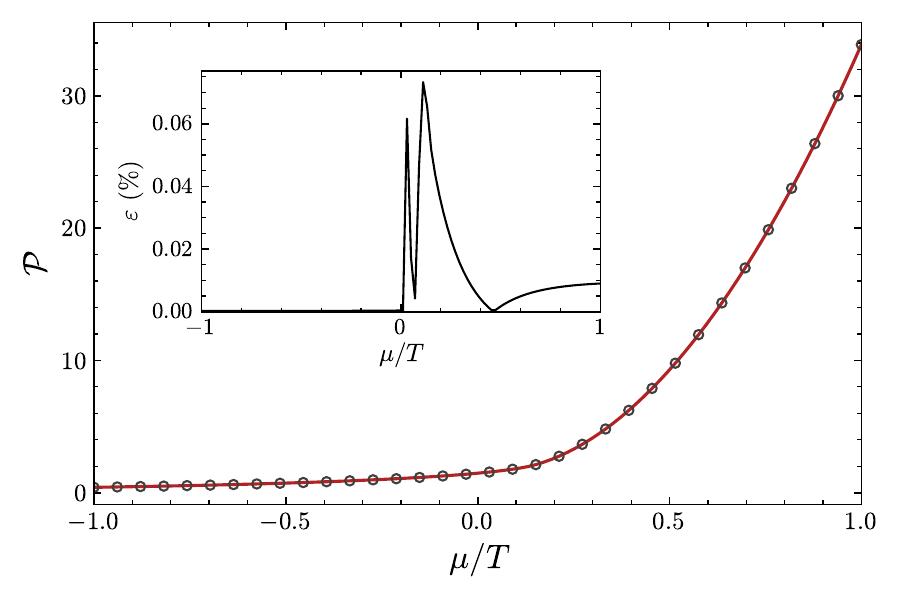}
\caption{Phase-space pressure $\mathcal P$ as a function of $\mu/T$ at $mg_0=0.1$, obtained within LPA$'$ (Wetterich's regulator) with (circles) and without (solid red line) the term $Z_{\tau,k}|\partial_\tau\phi|^2$ in Eq.~\eqref{eq:terme_Ztau}. The inset shows the relative difference $\varepsilon$, which does not exceed $0.1\,\%$.}
\label{fig:pression_Ztau}
\end{figure}

We also check within LPA$'$ that the second-order time-derivative term $Z_{\tau,k}|\partial_\tau\phi|^2$, Eq.~\eqref{eq:terme_Ztau}, omitted from our ansatz, has a negligible effect on thermodynamics.  Figure~\ref{fig:pression_Ztau} compares the phase-space pressure obtained with and without this term. The relative difference does not exceed $0.1\,\%$ on either side of the BKT transition and vanishes exactly for $\mu\leq0$.

These results indicate that the equation of state is remarkably insensitive to the approximation scheme, even at strong coupling. In particular, the field-expanded LPA$'$ already provides a quantitatively accurate description of the pressure, and the additional field dependence and spatial derivative terms introduce only small corrections. This suggests that the thermodynamics is well captured by relatively simple FRG approximations.

The situation is different for the superfluid density and the BKT transition. Within LPA$'$, the beta function of the superfluid density, $k\partial_k n_{s,k}$, remains small but never vanishes in the low-temperature phase. The flow follows a line of quasi-fixed points without reaching a true fixed point, and the superfluid density ultimately vanishes at a finite scale. In practice, the flow must be stopped at an arbitrary scale $k_{\rm min}$, leading to a superfluid density that depends on this choice. This limitation becomes more pronounced at strong coupling~\cite{ranconUniversalThermodynamicsTwodimensional2012}.

\begin{figure}[]
\centering
\includegraphics[width=0.6\textwidth]{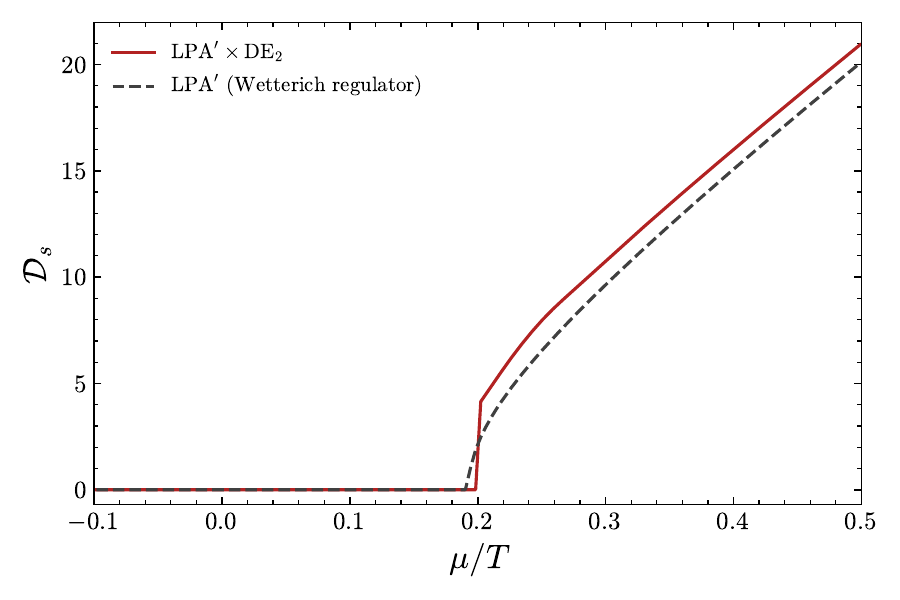}
\caption{Superfluid phase-space density $\mathcal D_s$ as a function of $\mu/T$ at $mg_0=0.15$, obtained with LPA$'\times$DE$_2$ (solid line) and field-expanded LPA$'$ (dashed line). The flow is stopped at $t_{\rm min}=\ln(k_{\rm min}/\Lambda)=-12$, corresponding to $k_{\rm min}\simeq6\times10^{-5}$ for $\Lambda=10$. The LPA$'\times$DE$_2$ gives the universal jump $\mathcal D_s\simeq4$ at the BKT transition, whereas LPA$'$ gives a continuous vanishing of the superfluid density.
}
\label{fig:Ds_LPA_vs_DE2}
\end{figure}

Figure~\ref{fig:Ds_LPA_vs_DE2} compares the superfluid density obtained at $mg_0=0.15$. The LPA$'\times$DE$_2$ displays the expected jump at $\mu_c/T$, whereas the LPA$'$ gives a continuous superfluid density and does not describe the BKT transition satisfactorily. Thus, although the equation of state is already well described at the LPA$'$ level, the superfluid density is considerably more sensitive to the approximation. The LPA$'\times$DE$_2$ provides a substantial improvement in the description of the BKT transition, as discussed in detail in Sec.~\ref{sec:FRGandBKT}.

\section{Universal equation of state}
\label{sec:equation_etat_universelle}

\begin{figure}[]
\centering
\includegraphics[width=0.8\textwidth]{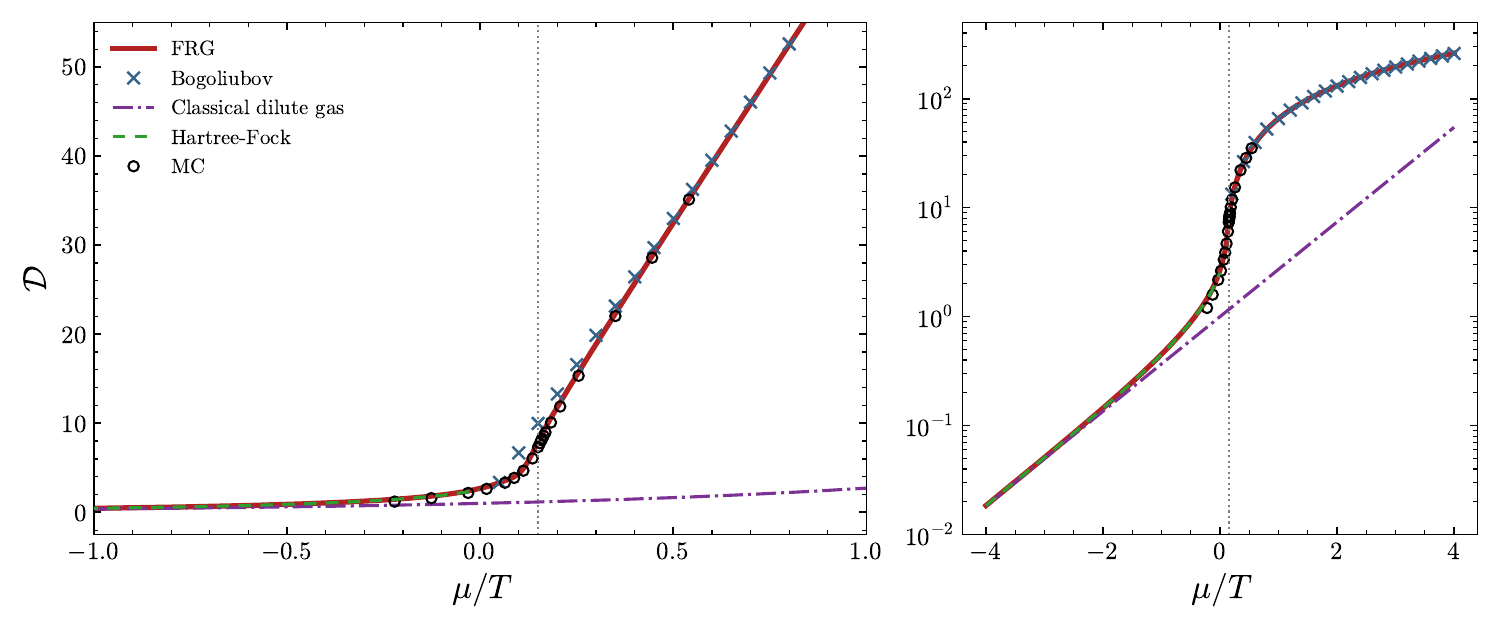}
\caption{Phase-space density $\mathcal D$ as a function of $\mu/T$ at $mg_0=0.1$, obtained from the FRG (red line) and compared with Monte Carlo data (open circles), Bogoliubov theory (crosses), self-consistent Hartree--Fock theory (dashed line), and the dilute classical gas (dot-dashed line). The vertical line marks the BKT transition.}
\label{fig:eos_comparaison}
\end{figure}

We now compute the universal equation of state of the dilute two-dimensional Bose gas. At fixed dimensionless coupling $\tilde g(T)$, we solve the flow equations for different values of $\mu/T$ to reconstruct $\mathcal D(\mu/T,\tilde g(T))$.
A typical result, obtained here for $mg_0=0.1$, is shown in Fig.~\ref{fig:eos_comparaison}.

For $\mu/T\ll-1$, the system is in the classical normal-fluid regime: thermal fluctuations dominate, and Boltzmann statistics applies. The phase-space density is
\begin{equation}
\mathcal{D}\!\left(\frac{\mu}{T}\right) = e^{\mu/T},
\end{equation}
corresponding to
\begin{equation}
\mathcal{F}(x,y) = e^x, \qquad x\ll-1,
\end{equation}
independently of the interaction $y=\tilde g(T)$. A better approximation is given by the Hartree-Fock approximation~\cite{hadzibabic_2011}, where the gas is treated as ideal with a chemical potential shifted by the mean interaction, $\mu_{\rm eff}=\mu-2 g_0 n$. This gives the implicit equation 
\begin{equation}
\mathcal{D}\!\left(\frac{\mu}{T}\right) =
-\ln\!\left(1-\exp\!\left(\frac{\mu}{T}
- \frac{\tilde{g}}{\pi}\mathcal{D}\right)\right).
\label{eq:HF}
\end{equation}

In the opposite limit, $\mu/T\gg1$ , the gas is quantum degenerate and thermal excitations become negligible. In weak coupling, it is then well described by Bogoliubov theory, with density~\cite{schick_1971,Mora09}
\begin{equation}
\mathcal{D}\!\left(\frac{\mu}{T},\tilde{g}(\mu)\right) =
\frac{\mu}{T}\left(\frac{2\pi}{\tilde{g}(\mu)}
+ \frac{1}{2}\ln 2 - \frac{1}{2}\right),
\qquad \frac{\mu}{T}\gg1,
\label{eq:densite_limite_mu}
\end{equation}
where the coupling is evaluated at the dominant energy scale, set here by the chemical potential.

The FRG reproduces both asymptotic limits with excellent accuracy, as shown in Fig.~\ref{fig:eos_comparaison}. It also gives the full universal equation of state, including the vicinity of the BKT transition ($\mu/T\sim\mu_c/T$), where fluctuations are strongest and standard perturbative approaches become insufficient. In this regime, the thermodynamics can be captured via  Monte Carlo simulations of an effective classical field theory~\cite{Prokofev01,Prokofev02}, with very good agreement between our results and the Monte Carlo data. A detailed comparison between the two is given in Sec.~\ref{sec_MC}. 

The FRG combined with the derivative expansion therefore describes the different regimes of the dilute two-dimensional Bose gas within a single approximation, from the normal phase through the BKT transition to the superfluid phase. It requires neither a small interaction parameter nor a change of approximation between regimes. Conventional approaches are instead restricted to weak coupling or to a particular region of the phase diagram.

\begin{figure}[]
\centering
\includegraphics[width=0.75\textwidth]{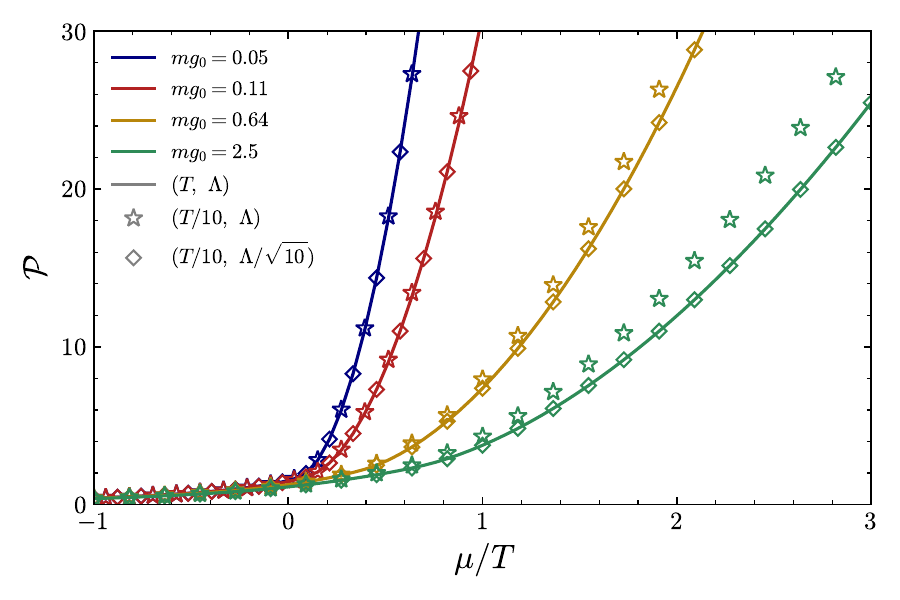}
\caption{Phase-space pressure $\mathcal P$ as a function of $\mu/T$ for several couplings $mg_0$ (colors) and three sets of parameters $(T,\Lambda)$ determining $\tilde g$: solid lines, $(T,\Lambda)$; stars, $(T/10,\Lambda)$; diamonds, $(T/10,\Lambda/\sqrt{10})$. }
\label{fig:invariance_echelle}
\end{figure}

We can also test the approximate scale invariance of the equation of state~\cite{Prokofev02,Yefsah11,Desbuquois14,pitaevskii_1997,Hung11}. Scale invariance holds in the weak-coupling limit $mg_0\ll1$, where $\tilde g(T)\simeq\tilde g=mg_0$, except at exponentially small temperatures, Eq.~\eqref{eq:couplage_g_epsilon}. At stronger coupling, the temperature dependence of $\tilde g(T)$ through the universal combination $ma_2^2T$ becomes significant. For fixed $mg_0$, this reduces to a dependence on $T/\Lambda^2$, as follows from Eq.~\eqref{eq:couplage_g_epsilon} and the cutoff dependence of $a_2$. Scale invariance is then broken.
This can be tested by computing the equation of state for different parameter sets~\cite{ranconUniversalThermodynamicsTwodimensional2012}. Replacing $(T,\Lambda)$ by $(T/10,\Lambda/\sqrt{10})$ leaves $T/\Lambda^2$, and hence $\tilde g(T)$, unchanged. The two sets are strictly equivalent for the scaling function. By contrast, $(T/10,\Lambda)$ changes $T/\Lambda^2$ and becomes equivalent to $(T,\Lambda)$ only at weak coupling.

Figure~\ref{fig:invariance_echelle} shows that scale invariance holds to an excellent approximation for $mg_0\lesssim0.11$, where the temperature dependence of $\tilde g$ at fixed $g_0$ is negligible. The three curves coincide, giving the same $\mathcal F(x,y)$ at fixed $y=\tilde g$. A simultaneous rescaling $\mu\to\lambda\mu$, $T\to\lambda T$ leaves $x=\mu/T$ unchanged and therefore leaves $\mathcal F(x,y)$ unchanged as long as $\tilde g(T)\simeq\tilde g(\lambda T)$. At stronger coupling, already for $mg_0=0.64$, the temperature dependence of $\tilde g$ breaks this scale invariance. At least two parameters must then be changed simultaneously to keep $\tilde g$ fixed, here $T$ and $\Lambda$. Thus $(T,\Lambda)$ and $(T/10,\Lambda/\sqrt{10})$ give the same $\mathcal F(x,y)$, whereas $(T/10,\Lambda)$ does not. 

We next characterize the BKT transition itself by extracting the critical temperature and the superfluid density at the transition.

\section{BKT transition and superfluid density}
\label{sec:FRGandBKT}

The FRG applied to the classical linear $O(2)$ model has been shown to capture most features of the BKT transition~\cite{grater_1995,gersdorff_2001,jakubczyk_2014}. Within this framework, the vortices are not introduced explicitly. At LPA$'$, one only finds a line of quasi-fixed-points, along which the RG flow is very slow. The system always flows to the disordered phase, albeit with an exponentially large correlation length at low temperature. One can nevertheless extract the critical temperature somewhat reliably from these flows, see e.g. \cite{Machado10,ranconUniversalThermodynamicsTwodimensional2012}. The failure of the LPA$'$ has been understood as coming from the mistreatment of the longitudinal fluctuations of the order parameter \cite{Jakubczyk17}. 

At DE$_2$, one does find a phase transition, with most universal features properly recovered \cite{jakubczyk_2014}.
In the high-temperature disordered phase, the running superfluid density $n_{s,k}$ and anomalous dimension $\eta_k$ vanish at a finite scale $k\sim\xi^{-1}$. The minimum of the effective potential also vanishes at this scale, with $\rho_{0,k}=0$ for $k\lesssim\xi^{-1}$. The correlation length extracted from the flow, $\xi=\lim_{k\to0}[Z_{x,k}(0)/U_k'(0)]^{1/2}$, reproduces the essential singularity characteristic of BKT theory~\cite{kosterlitz_1974,gersdorff_2001,jakubczyk_2014}.
In the low-temperature superfluid phase, the flow of the dimensionless minimum $\tilde\rho_{0,k}=Z_{x,k}\rho_{0,k}$ displays a quasi-fixed point for $\tilde\rho_{0,k}\geq\tilde\rho_{ 0,\star}$, defined by the near-vanishing of the beta function $\beta(\tilde\rho_{0,k})=\partial_t\tilde\rho_{0,k}\simeq0$. Here $\tilde\rho_{0,\star}$ is the critical value of the dimensionless minimum: below it, $\tilde\rho_{0,k}$ vanishes at finite $k$ (disordered phase); above it, the flow develops a quasi-plateau approaching a finite value as $k\to0$ (superfluid phase). The quasi-fixed point can be made exact by optimizing the nonphysical regulator parameter $\alpha$. This yields a line of fixed points for $T<T_{\rm BKT}$, in agreement with the BKT scenario~\cite{jakubczyk_2014}. Thus $n_{s,k}$ and $\eta_k$ approach finite plateaus as $k\to0$, whereas $\rho_{0,k}$ vanishes as $k^\eta$, Fig.~\ref{fig:flot_boson_UZeta}. This behavior is consistent with the quasi-long-range order and infinite correlation length of the BKT phase.

The two phases are separated by the transition at $T_{\rm BKT}$. It can be located directly from the limiting behavior of $n_{s,k}$ and $\eta_k$, giving access to the universal values $\eta(T_{\rm BKT})=1/4$ and $\mathcal D_s(T_{\rm BKT})=4$~\cite{kosterlitz_1974,nelson_1977}. These results for the classical $O(2)$ model provide a reference for the dilute two-dimensional Bose gas. We expect the same behavior in the effective classical regime $k\lesssim k_{\rm dB}$, where the gas reduces to a classical $O(2)$ model.

\begin{figure}[t]
\centering
\includegraphics[width=0.55\textwidth]{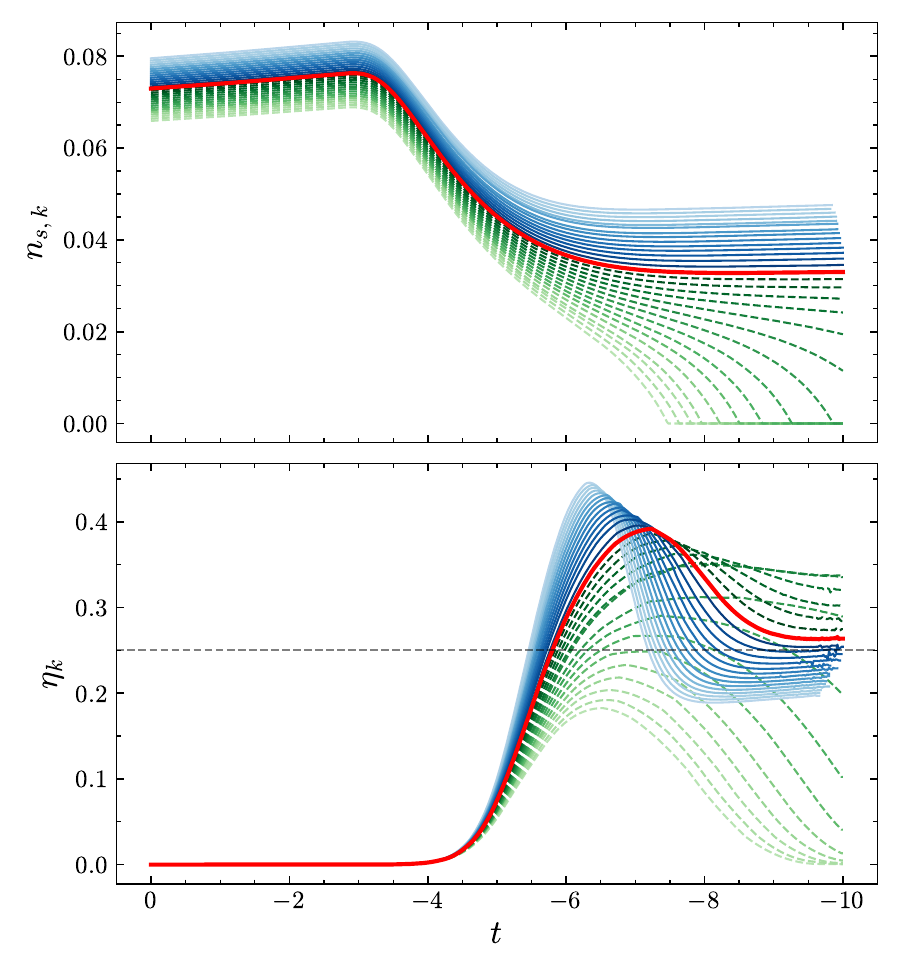}
\caption{RG flow of the running superfluid density $n_{s,k}$ (top) and anomalous dimension $\eta_k$ (bottom) as functions of RG time $t$, for several initial chemical potentials $\mu$ near $\mu_c$, at $mg_0=0.11$. Green dashed curves correspond to $\mu<\mu_c$ (normal phase), and blue solid curves to $\mu>\mu_c$ (quasi-long-range order phase). Darker colors represent trajectories closer to the BKT point. The critical trajectory $\mu=\mu_c$ is shown in red and develops a quasi-plateau. In the lower panel, the horizontal dashed line marks the universal BKT value $\eta=1/4$.}
\label{fig:flot_ns_eta_t}
\end{figure}

In the effective classical regime $k\lesssim k_{\rm dB}$, the BKT transition is fully determined by the structure of the RG flow. The initial condition, set by $\mu/T$ at fixed $\tilde g$, determines whether the trajectory approaches the fixed-point line of the effective classical $O(2)$ model or moves away from it toward the massive regime. Figure~\ref{fig:flot_ns_eta_t} shows the flows of $\eta_k$ and $n_{s,k}$ for several values of $\mu/T$ near the transition.

For $\mu/T<\mu_c/T$, the flow leaves the fixed-point line and enters the massive regime, where $n_{s,k}$ and $\eta_k$ vanish at a finite scale $k\sim\xi^{-1}$. The effective-potential minimum vanishes at the same scale, $\rho_{0,k}=0$ for $k\lesssim\xi^{-1}$, characterizing the normal phase. Conversely, for $\mu/T>\mu_c/T$, the flow approaches the fixed-point line, and $n_{s,k}$ and $\eta_k$ remain finite as $k\to0$, developing quasi-plateaus. The minimum $\rho_{0,k}$ then vanishes only asymptotically, following the power law $\rho_{0,k}\propto k^\eta$ characteristic of the superfluid phase, as shown in Fig.~\ref{fig:flot_boson_UZeta}.

The critical trajectory $\mu/T=\mu_c/T$ is the separatrix between these two behaviors. Along this trajectory, $\eta_k$ and $n_{s,k}$ reach plateaus whose limiting values determine the critical ratio $\mu_c/T$, the universal jump $\mathcal D_s(T_{\rm BKT})$, and the anomalous dimension $\eta(T_{\rm BKT})$. For $mg_0=0.11$, we obtain
\begin{equation}
\begin{aligned}
\mu_c/T &= 0.161\pm 0.006, \\
\mathcal{D}_s(\mu_c^+/T) &= 4.15\pm 0.17, \\
\eta(\mu_c^+/T) &= 0.26\pm 0.02,
\end{aligned}
\end{equation}
in good agreement with the expected universal values $\mathcal D_s(\mu_c^+/T)=4$ and $\eta(\mu_c^+/T)=1/4$. The error bars are estimated numerically from the sensitivity to the integration scheme and to the extraction of the asymptotic plateau as $k\to0$.

\begin{figure}[]
\centering
\includegraphics[width=0.65\textwidth]{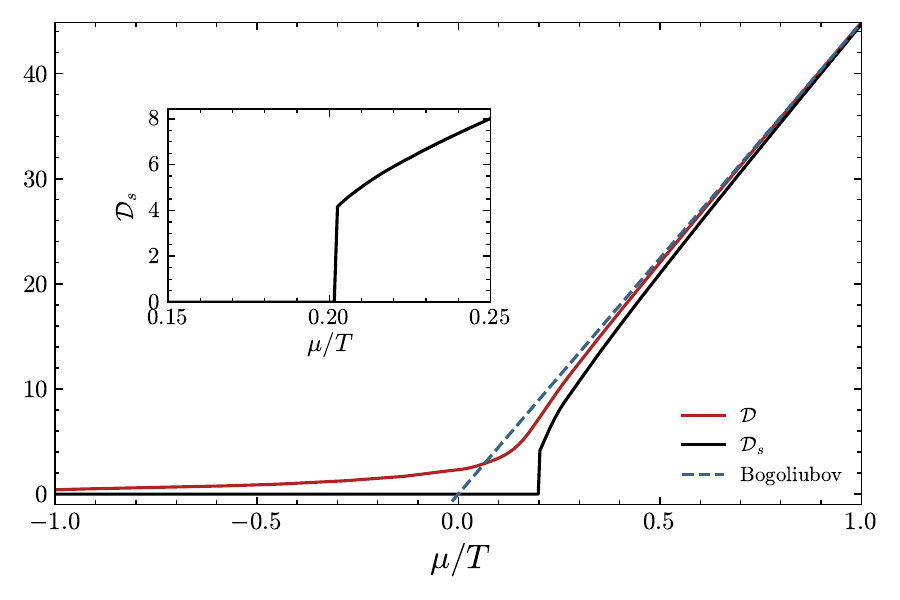}
\caption{Total phase-space density $\mathcal D$ and superfluid phase-space density $\mathcal D_s$ as functions of $\mu/T$ at $mg_0=0.15$. The superfluid density vanishes in the normal phase, below $\mu_c/T\approx0.202$. At the BKT transition, $\mathcal D_s$ has a universal discontinuous jump of height $4.1$, close to the universal BKT value of $4$, whereas $\mathcal D$ remains smooth throughout. The inset shows the superfluid-density jump at the transition. The Bogoliubov phase-space density in the degenerate regime is also shown.  }
\label{fig:PlotDandDs}
\end{figure}

As shown in Fig.~\ref{fig:PlotDandDs}, the total density $\mathcal D$ remains smooth across the BKT transition, whereas $\mathcal D_s$ has the expected discontinuous jump. In the zero-temperature limit, $\mathcal D_s\to\mathcal D$: the gas is entirely superfluid. The Ward identity associated with Galilean invariance imposes $Z_{x,k=0}(\rho_{0,k=0})=n/\rho_{0,k=0}$ at $T=0$, and hence $n_s=n=-\lim_{k\to0}\partial_\mu U_k(\rho_{0,k})$~\cite{gavoret_1964,pistolesi_2004,dupuis_2009a}. 

\begin{figure}[th!]
\centering
\includegraphics[width=0.7\textwidth]{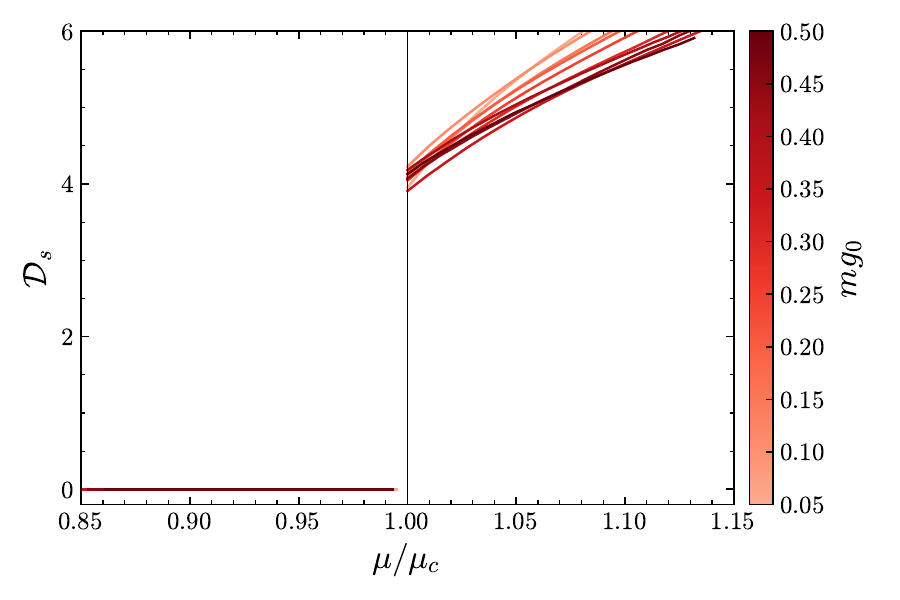}
\caption{Superfluid phase-space density $\mathcal D_s$ as a function of $\mu/\mu_c$ for several couplings $mg_0$, indicated by the color scale. It vanishes in the normal phase, $\mu/\mu_c<1$, and has a universal discontinuous jump close to $4$ at the BKT transition, $\mu/\mu_c=1$, independently of $m g_0$.}
\label{fig:Ds_mu_muc_multicoupling}
\end{figure}

Varying the interaction $g_0$, we also verify that the universal jump $\mathcal D_s(T_{\rm BKT})$ is independent of $mg_0$, although superfluidity itself requires interactions. Figure~\ref{fig:Ds_mu_muc_multicoupling} shows that the FRG reproduces $\mathcal D_s(T_{\rm BKT})=4$ to within $5\,\%$ over the entire coupling range studied, even beyond the weakly interacting regime.

\section{Comparison to classical field simulations}
\label{sec_MC}

We now compare our results to classical-field Monte Carlo calculations of
Refs.~\cite{Prokofev01,Prokofev02}. For $\tilde g\equiv\tilde g(T)\ll1$ and $|\mu|\ll T$, the
interaction scale $k_c\sim\sqrt{mT\tilde g}$ is much smaller than the thermal
scale $k_T\sim\sqrt{mT}$. The strongly fluctuating modes are therefore
classical and described by an effective classical $|\psi|^4$ theory: one can integrate out the quantum modes (with finite Matsubara frequencies) and classical modes at scales smaller than the de Broglie thermal length~\cite{popov_1983,kagan_1987,fisher_1988,Prokofev01}. While weakly coupled, this classical field theory has strong fluctuations at large scales nevertheless. It cannot be studied perturbatively, and Refs.~\cite{Prokofev01,Prokofev02} solved this issue by performing classical Monte Carlo simulations of the effective classical field theory.

One nevertheless needs to relate the classical observables to the quantum ones by matching the two theories at some intermediate scale. Note that this whole procedure is only valid at weak coupling, and is known to give unphysical results (e.g. a superfluid density larger than the density) too far from the fluctuation regime already at interactions of order $mg_0\simeq 0.2$. In the FRG framework, this procedure is performed automatically along the RG flow, as the effect of the finite Matsubara modes is included in the long-distance physics. Furthermore, they decouple from the flow for $k\lesssim \sqrt{mT}$, leaving us with an effective classical field theory. The advantage here is that the procedure does not rely on weak coupling.

It has been shown in~\cite{Prokofev01,Prokofev02} that the critical chemical potential $\mu_c$ and the critical density $n_c=n(\mu_c,T)$ take the universal forms
\begin{align}
\frac{\mu_c}{T} &= \frac{\tilde{g}}{\pi}
\ln\left(\frac{\xi_\mu}{\tilde{g}}\right) \label{eq:muc_classicalfield} \\
\lambda_{\rm dB}^{2}n_c &= 
\ln\left(\frac{\xi_n}{\tilde{g}}\right) \label{eq:nc_classicalfield}
\end{align}
where $\xi_\mu=13.2\pm0.4$ and $\xi_n=380\pm3$ encode the nonperturbative physics of the critical region. The logarithmic form of these relations arises from the Gaussian behavior of the high-momentum modes. Only the universal constants $\xi_\mu$ and $\xi_n$ must be determined nonperturbatively. Thus Eqs.~\eqref{eq:muc_classicalfield} and~\eqref{eq:nc_classicalfield} are universal functions of $\tilde g$. These forms were subsequently confirmed by independent quantum Monte Carlo calculations~\cite{pilati_2008,holzmann_2008}.

\begin{figure}
  \centering
\includegraphics[width=0.85\textwidth]{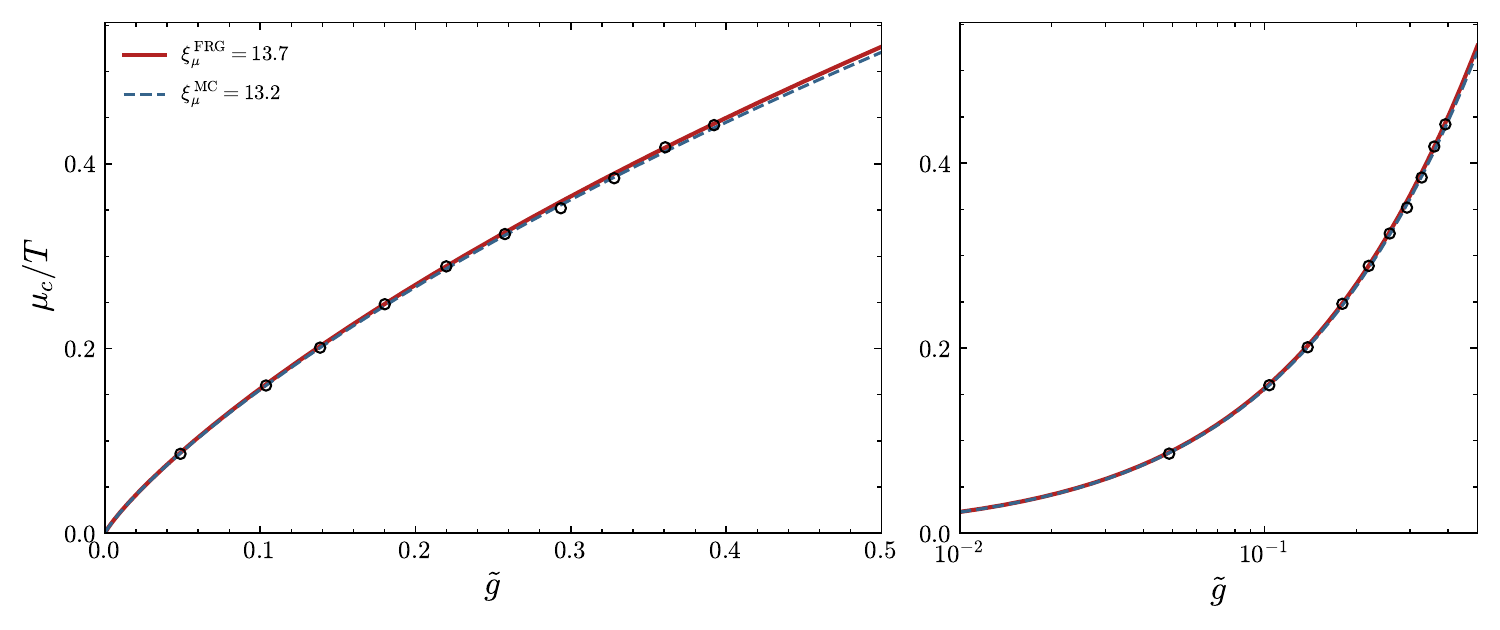}
\caption{Critical chemical potential $\mu_c/T$ as a function of the dimensionless coupling $\tilde g$, on a linear scale (left) and on a logarithmic scale (right). Open circles are the values extracted from the FRG flow combined with the LPA$'\times$DE$_2$. The solid red curve is a fit of the universal form~\eqref{eq:muc_classicalfield} to these points, yielding $\xi_\mu^{\rm FRG}=13.7$, to be compared with the dashed blue curve, corresponding to the Monte Carlo value $\xi_\mu^{\rm MC}=13.2$~\cite{Prokofev01}.}
\label{fig:muc_comparaison}
\end{figure}

\begin{figure}
\centering
\includegraphics[width=0.95\textwidth]{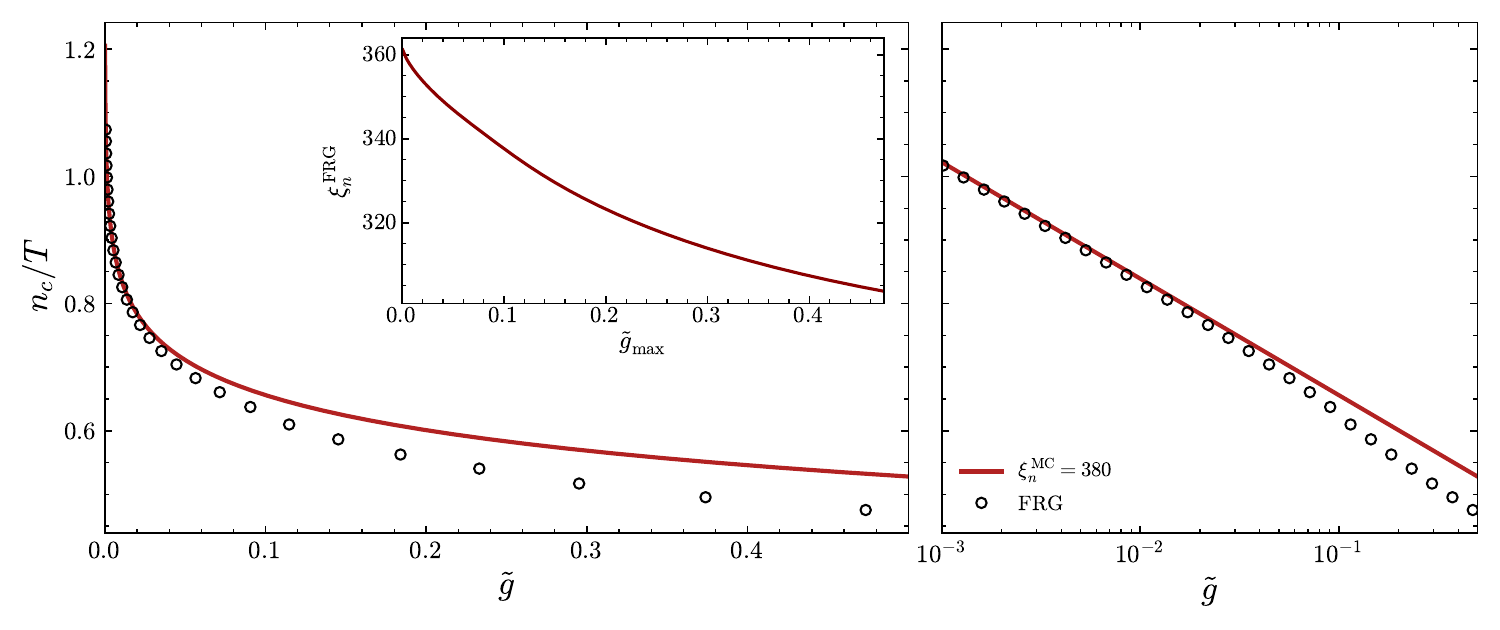}
\caption{Critical density $n_c/T$ as a function of the dimensionless coupling $\tilde g$ in linear (left) and semi-log (right) scales, in the weak-interaction regime. The points are obtained with the FRG at the LPA$'\times$DE$_2$ level, and the red line corresponds to the universal law, Eq.~\eqref{eq:nc_classicalfield}, with $\xi_n^{\rm MC}=380$. The inset shows the constant $\xi_n^{\rm FRG}$ obtained by fitting the universal law over the interval $\tilde g\leq\tilde g_{\max}$, as a function of $\tilde g_{\max}$.}
\label{fig:nc_T_vs_g_FRG_MC}
\end{figure}

Within the FRG, we extract $\mu_c/T$ from the jump of the superfluid density and $n_c$ from the density, and fit the universal constants $\xi_\mu$ and $\xi_n$. The results are shown in Figs.~\ref{fig:muc_comparaison} and~\ref{fig:nc_T_vs_g_FRG_MC}.

Regarding $\mu_c/T$, the agreement between the FRG and classical field simulations is excellent. The fit of the FRG data yields 
\begin{equation}
\xi_\mu^{\rm FRG}=13.7\pm1.4,
\label{eq:xi_mu_FRG_fit}
\end{equation}
a value consistent, within error bars, with $\xi_\mu^{\rm MC}=13.2\pm0.4$. This is to be contrasted with the estimate obtained within the field-expanded LPA$'$ calculation giving $\xi_\mu^{\rm LPA'}\simeq 9.5$.
The logarithmic dependence on $\tilde g$ persists up to $\tilde g\sim1$, well beyond the weak-interaction regime $\tilde g\ll1$, confirming that the universal scaling law of Eq.~\eqref{eq:muc_classicalfield} remains quantitatively accurate up to the strong-coupling regime.

The critical density obtained from the FRG also agrees well with the universal prediction of Eq.~\eqref{eq:nc_classicalfield} at weak coupling, using the classical-field value $\xi_n^{\mathrm{MC}}=380$ (Fig.~\ref{fig:nc_T_vs_g_FRG_MC}). Unlike the critical chemical potential, however, the agreement deteriorates with increasing $\tilde g$, with deviations of about $3\,\%$ at $\tilde g\simeq 0.1$. The value of $\xi_n$ extracted from a fit to Eq.~\eqref{eq:nc_classicalfield} therefore depends on the coupling range considered (see the inset of Fig.~\ref{fig:nc_T_vs_g_FRG_MC}). It is only for $\tilde g\lesssim 10^{-2}$ that the estimate stabilizes to
\begin{equation}
\xi_n^{\rm FRG} = 361.9\pm0.9.
\label{eq:xi_n_FRG_fit}
\end{equation}
This value is about $5\,\%$ smaller than the classical-field estimate, $\xi_n^{\rm MC}=380\pm3$~\cite{Prokofev01}.

A summary of these results is given in Table~\ref{tab:xi_n_xi_mu_comparaison}, where we also report the results obtained from complex Langevin simulations \cite{heinen_2022,heinen_2023}. Our results are in better agreement with those of classical field simulations than with the latter. 

\begin{table}[]
\centering
\begin{tabular}{lcc}
\toprule
 & $\xi_n$ & $\xi_\mu$ \\
\midrule
MC & $380 \pm 3$ & $13.2 \pm 0.4$ \\
CL & $260 \pm 12$ & $12.3 \pm 0.1$ \\
LPA$'\times$DE$_2$ & $361.9 \pm 0.9$ & $13.7 \pm 1.4$ \\
\bottomrule
\end{tabular}
\caption{Universal constants $\xi_n$ and $\xi_\mu$ obtained from classical-field Monte Carlo simulations (MC)~\cite{Prokofev01}, complex Langevin simulations (CL)~\cite{heinen_2023}, and the FRG at the LPA$'\times$DE$_2$ level.}
\label{tab:xi_n_xi_mu_comparaison}
\end{table}

\begin{figure}[t]
\centering
\includegraphics[width=0.6\textwidth]{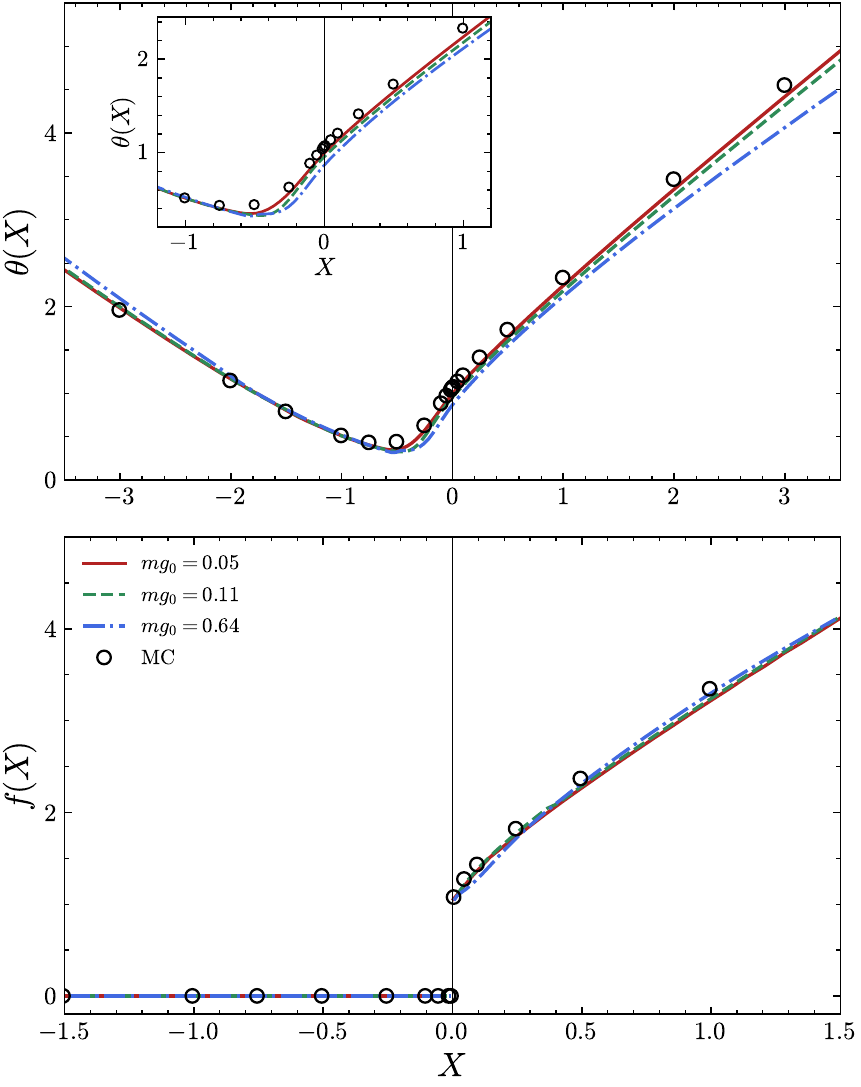}
\caption{Functions $\theta(X)$ (top) and $f(X)=\mathcal D_s/4$ (bottom) as functions of the control parameter $X$, obtained from the FRG within the LPA$'\times$DE$_2$ at several couplings $mg_0$ and compared with the Monte Carlo simulations of Ref.~\cite{Prokofev02} (open circles). Inset: Zoom on the fluctuation regime $|X|\lesssim1$. Same legend for both panels.}
\label{fig:theta_X_FRG_MC}
\end{figure}

In \cite{Prokofev02}, the equation of state for the density was given in terms of the universal function
\begin{equation}
\theta(X) \equiv \frac{2n}{mT} - \frac{\mu}{T\tilde{g}(T)},
\label{eq:theta_definition_finale}
\end{equation}
with control parameter
\begin{equation}
X = \frac{\mu-\mu_c}{\tilde g(T)T}.
\label{eq:X_definition_finale}
\end{equation}
This form is a priori valid in the fluctuation regime, where $X$ is of order unity, and at weak coupling. The phase-space density can then be expressed in terms of $\theta(X)$,
\begin{equation}
\mathcal D(\mu,T) = \mathcal D_c
+ \pi\big[\theta(X) - \theta_0 + X\big],
\label{eq:D_via_theta_finale}
\end{equation}
with $\theta_0=\theta(0)$, and assuming Eqs.~\eqref{eq:muc_classicalfield} and \eqref{eq:nc_classicalfield}, $\theta_0=\pi^{-1}\ln(\xi_n/\xi_\mu)$. 
The function $\theta(X)$ was tabulated from classical simulations in \cite{Prokofev02}. 
Furthermore, the superfluid density as a function of $X$ was also tabulated in that reference, in terms of a function $f=\mathcal{D}_s/4$, for which the superfluid jump is one.
Figure~\ref{fig:theta_X_FRG_MC} shows these two scaling functions compared to our FRG results. Concerning the superfluid density (lower panel), the comparison is very good, and we observe that the FRG results do not depend on $mg_0$. Concerning the scaling function $\theta$ (upper panel), we observe that while the overall shape is similar between classical field simulations and FRG, the FRG results have a clear dependence on $mg_0$, especially visible at larger $mg_0$. First, there is a downward shift around $X=0$ as $mg_0$ increases, due to the dependence of $n_c$ on $\tilde g$ beyond Eq.~\eqref{eq:nc_classicalfield}. 
Furthermore, away from the fluctuation regime, the energy dependence of the renormalized interaction becomes important, as exemplified by the Bogoliubov result \eqref{eq:densite_limite_mu} in the degenerate regime. This effect is more pronounced at stronger coupling. 
Deviations from classical-field predictions have also been observed in measurements of Tan's contact at comparable interaction strengths, with which the FRG results are in better agreement \cite{Rancon23}.
These deviations originate mainly from the classical-field estimate of the density and can thus be traced back to the deviations in the scaling function $\theta(X)$ discussed above.

\section{Conclusion \label{sec_concl}}

We have computed the universal thermodynamics of the dilute two-dimensional Bose gas using the functional renormalization group, retaining the full field dependence of the effective potential and spatial derivative terms. This allows us to obtain the equation of state and the superfluid density across the BKT transition, for both weak and strong interactions.

We find that the equation of state is remarkably insensitive to the approximation scheme. The simpler LPA$'$ and the LPA$'\times$DE$_2$ give pressures that differ by only a few percent, even at the largest couplings considered. The situation is different for the superfluid density. While the LPA$'$ fails to describe the BKT transition correctly, the LPA$'\times$DE$_2$ reproduces the universal jump of the superfluid density and the critical anomalous dimension to within a few percent. 

At weak coupling, our results are in very good agreement with classical-field Monte Carlo simulations, both for the critical parameters and for the universal scaling function of the density in the fluctuation regime. At stronger coupling, we observe deviations from the classical-field scaling, while the more general quantum scaling form remains applicable. Our results thus provide predictions for the thermodynamics of planar Bose gases beyond the weak-coupling regime, where classical-field approaches are no longer quantitatively reliable.

The simultaneous determination of the equation of state and the superfluid density also opens the possibility of computing observables that were previously out of reach within FRG approaches. In particular, these quantities provide the necessary input for calculating the velocities of first and second sound within two-fluid hydrodynamics \cite{ozawa_2014,ota_2018}. These collective modes are accessible in ultracold-atom experiments at strong coupling \cite{Christodoulou21}.

\acknowledgments
We thank Nicolas Dupuis for many discussions on this topic over the years.
A.R. and I.B. are supported by the project HRZZ-IP-10-2022-9423. I.B. acknowledges support from the project “Implementation of cutting-edge research and its application as part of the Scientific Center of Excellence for Quantum and Complex Systems, and Representations of Lie Algebras”, Grant No. PK.1.1.10.0004, co-financed by the European Union through the European Regional Development Fund - Competitiveness and Cohesion Programme 2021-2027, and the support of the INFaR and FrustKor projects financed by the EU through the National Recovery and Resilience Plan (NRRP) 2021-2026 and the KaCIF project, co-financed by the Croatian Government and the European Union through the European Regional Development Fund---Competitiveness and Cohesion Operational Program (Grant No. KK.01.1.1.02.0012). A.R. has benefited from the financial support of the Grant No. ANR-24-CE30-6695 FUSIoN, ANR-24-CE47-4949 UniQ-RingS, ANR-23-PETQ-0001 Dyn1D at the title of France 2030, and Labex CEMPI (Grant No. ANR-11-LABX-0007-01).

\appendix

\section{FRG flow equations \label{sec_Appendix1}}

In this appendix, we give the FRG flow equations of the dilute Bose gas in dimension $d$ and at finite temperature $T$ at second order of the derivative expansion (DE$_2$), obtained from the Wetterich equation~\eqref{eq:flot_wetterich_boson}. We then give the flow equations used in the remainder of the paper, i.e. at LPA$'\times$DE$_2$. We consider the most general ansatz compatible with the symmetries of the action~\eqref{eq_action_boson} at second order in derivatives, in which all the coefficients retain their full functional dependence on $\rho$,
\begin{equation}
\begin{aligned}
\Gamma_{k}[\phi^*,\phi] = \int_0^\beta d\tau \int d^d r \Big\{
&Z_{\tau1,k}(\rho)\,\phi^*\partial_\tau\phi
+Z_{\tau,k}(\rho)\left|\partial_\tau\phi\right|^2
+Z_{x,k}(\rho)\left|\nabla\phi\right|^2 \\
&+\frac{Y_{\tau,k}(\rho)}{4}\left(\partial_\tau\rho\right)^2
+\frac{Y_{x,k}(\rho)}{4}\left(\nabla\rho\right)^2
+U_k(\rho)\Big\}.
\end{aligned}
\label{eq:ansatz_DE2_annexe}
\end{equation}
The propagators are obtained from this ansatz according to Eq.~\eqref{eq:Gk_definition}, and the flows of the various coefficients are obtained by projecting the flow of $\Gamma_k^{(2)}$,  Eq.~\eqref{eq:flot_Gamma2}, as described in Sec.~\ref{sec_FRG}.

In what follows, the operator $\tilde\partial_t$ is defined in Eq.~\eqref{eq:dt_tilde} and acts only on the $k$ dependence of the regulator $R_k$. The Matsubara
frequencies are $\omega_n=2\pi nT$. To aleviate the notation, we omit in
the integrands the index $k$ as well as the arguments of the coefficients
and propagators, $G_{ij}\equiv G_{ij,k}(\bm p,\omega_n;\rho)$ with
$ij\in\{LL,TT,LT\}$, $Z_x\equiv Z_{x,k}(\rho)$, etc. A prime on the
coefficients denotes the derivative with respect to $\rho$, and we write
$V_k(\rho)=\partial_\rho U_k(\rho)$, so that
$V'=\partial_\rho^2U_k$ and $V''=\partial_\rho^3U_k$. The flow
equations then read
\begin{align}
&\partial_t V_k(\rho) ={} - \frac{1}{2}\,\tilde\partial_t\,\frac{1}{\beta}\sum_{\omega_n}\int_{\bm p}\,\Big\{ G_{LL}\Big(V' + 2\,\rho V'' + \bm p^2 Z_x' + \omega_n^{2} Z_\tau' + \rho \bm p^2 Y_x' + \rho \omega_n^{2} Y_\tau'\Big)
\notag\\
&\qquad - 2 \omega_n\,G_{LT}\big(2\,Z_{\tau1}' + \rho Z_{\tau1}''\big) + G_{TT}\Big(3\,V' + \bm p^2 Z_x' + \bm p^2 Y_x + \omega_n^{2} Z_\tau' + \omega_n^{2} Y_\tau\Big)
\notag\\
&\qquad + \rho\,G_{LL} G_{TT}\big(2\,V' + \bm p^2 Y_x + \omega_n^{2} Y_\tau\big)^{2} + \rho\,G_{LT}^{2}\big(2\,V' + \bm p^2 Y_x + \omega_n^{2} Y_\tau\big)^{2}\Big\},
\label{eq:flot_V_DE2}
\end{align}

\begin{align}
&\partial_t Z_{x,k}(\rho) ={} -\frac{1}{2d}\,\tilde\partial_t\,\frac{1}{\beta}\sum_{\omega_n}\int_{\bm p}\,\Big\{ d\,G_{LL}\Big(Z_x' + Y_x + 2\,\rho Z_x'' + 5\,\rho Y_x' + 2\,\rho^{2} Y_x''\Big)
\notag\\
&\qquad + d\,G_{TT}\big(Z_x' + \rho Y_x'\big) + 2 \rho\,G_{LL}^{2}\big(Z_x' + Y_x + \rho Y_x'\big)\Big(6\,d V' + \bm p^2 Z_x' + \bm p^2 Y_x + 4\,d \rho V''
\notag\\
&\qquad\quad + 2\,d \bm p^2 Z_x' + 2\,d \bm p^2 Y_x + \rho \bm p^2 Y_x' + 2\,d \omega_n^{2} Z_\tau' + 2\,d \omega_n^{2} Y_\tau + 2\,d \rho \bm p^2 Y_x' + 2\,d \rho \omega_n^{2} Y_\tau'\Big)
\notag\\
&\qquad - 8 d \rho \omega_n\,G_{LL} G_{LT}\big(2\,Z_{\tau1}' + \rho Z_{\tau1}''\big)\big(Z_x' + Y_x + \rho Y_x'\big)
\notag\\
&\qquad - 2 \rho\,G_{LT}^{2}\Big(2\,\bm p^2 Z_x'^{2} + 2\,d V' Z_x' + 5\,d V' Y_x + 2\,\bm p^2 Z_x' Y_x + 2\,d \bm p^2 Z_x'^{2} + d \bm p^2 Y_x^{2} + 2\,d \rho V' Y_x'
\notag\\
&\qquad\quad + 2\,d \rho V'' Y_x + 3\,d \bm p^2 Z_x' Y_x + 2\,\rho \bm p^2 Z_x' Y_x' + 2\,d \omega_n^{2} Z_x' Z_\tau' + 3\,d \omega_n^{2} Y_x Z_\tau' + d \omega_n^{2} Y_x Y_\tau
\notag\\
&\qquad\quad + 2\,d \rho \bm p^2 Z_x' Y_x' + d \rho \bm p^2 Y_x Y_x' + d \rho \omega_n^{2} Y_x Y_\tau' + 2\,d \rho \omega_n^{2} Y_x' Z_\tau'\Big)
\notag\\
&\qquad - 4 d \rho \omega_n Y_x\,G_{LT} G_{TT}\big(2\,Z_{\tau1}' + \rho Z_{\tau1}''\big) + 2 \rho\,G_{TT}^{2}\Big(\bm p^2 Z_x'^{2} + d V' Y_x + d \bm p^2 Z_x' Y_x + d \omega_n^{2} Y_x Z_\tau'\Big)
\notag\\
&\qquad - 4 \rho \bm p^2\,\big(\partial_{\bm p^2}G_{LL}\big)^{2}\Big(3\,V' + 2\,\rho V'' + \bm p^2 Z_x' + \bm p^2 Y_x + \omega_n^{2} Z_\tau' + \omega_n^{2} Y_\tau + \rho \bm p^2 Y_x' + \rho \omega_n^{2} Y_\tau'\Big)^{2}
\notag\\
&\qquad + 8 \rho \bm p^2\,\big(\partial_{\bm p^2}G_{LT}\big)^{2}\Big(3\,V'^{2} + 2\,\rho V' V'' + 4\,\bm p^2 V' Z_x' + \bm p^2 V' Y_x + \bm p^{4} Z_x'^{2} - 4\,\omega_n^{2} Z_{\tau1}'^{2} + \bm p^{4} Z_x' Y_x
\notag\\
&\qquad\quad + 4\,\omega_n^{2} V' Z_\tau' + \omega_n^{2} V' Y_\tau + \rho \bm p^2 V' Y_x' + 2\,\rho \bm p^2 V'' Z_x' + \rho \bm p^{4} Z_x' Y_x' + \rho \omega_n^{2} V' Y_\tau' + 2\,\rho \omega_n^{2} V'' Z_\tau'
\notag\\
&\qquad\quad - 4\,\rho \omega_n^{2} Z_{\tau1}' Z_{\tau1}'' + 2\,\bm p^2 \omega_n^{2} Z_x' Z_\tau' + \bm p^2 \omega_n^{2} Z_x' Y_\tau + \bm p^2 \omega_n^{2} Y_x Z_\tau' + \omega_n^{4} Z_\tau'^{2} - \rho^{2} \omega_n^{2} Z_{\tau1}''^{2} + \omega_n^{4} Z_\tau' Y_\tau
\notag\\
&\qquad\quad + \rho \bm p^2 \omega_n^{2} Z_x' Y_\tau' + \rho \bm p^2 \omega_n^{2} Y_x' Z_\tau' + \rho \omega_n^{4} Z_\tau' Y_\tau'\Big) - 4 \rho \bm p^2\,\big(\partial_{\bm p^2}G_{TT}\big)^{2}\big(V' + \bm p^2 Z_x' + \omega_n^{2} Z_\tau'\big)^{2}
\notag\\
&\qquad + 16 \rho \bm p^2 \omega_n\,\partial_{\bm p^2}G_{LL} \partial_{\bm p^2}G_{LT}\big(2\,Z_{\tau1}' + \rho Z_{\tau1}''\big)\Big(3\,V' + 2\,\rho V'' + \bm p^2 Z_x' + \bm p^2 Y_x + \omega_n^{2} Z_\tau' + \omega_n^{2} Y_\tau
\notag\\
&\qquad\quad + \rho \bm p^2 Y_x' + \rho \omega_n^{2} Y_\tau'\Big) + 8 \rho \bm p^2 \omega_n^{2}\,\partial_{\bm p^2}G_{LL} \partial_{\bm p^2}G_{TT}\big(2\,Z_{\tau1}' + \rho Z_{\tau1}''\big)^{2}
\notag\\
&\qquad + 16 \rho \bm p^2 \omega_n\,\partial_{\bm p^2}G_{LT} \partial_{\bm p^2}G_{TT}\big(2\,Z_{\tau1}' + \rho Z_{\tau1}''\big)\big(V' + \bm p^2 Z_x' + \omega_n^{2} Z_\tau'\big)\Big\},
\label{eq:flot_Zx_DE2}
\end{align}

\begin{align}
&\partial_t \big[Z_{x,k}(\rho)+\rho\,Y_{x,k}(\rho)\big] ={} -\frac{1}{2d}\,\tilde\partial_t\,\frac{1}{\beta}\sum_{\omega_n}\int_{\bm p}\,\Big\{ d\,G_{LL}\big(Z_x' + 2\,\rho Z_x''\big) + d\,G_{TT}\big(Z_x' + Y_x\big)
\notag\\
&\qquad + \rho\,G_{LL} G_{TT}\Big(4\,\bm p^2 Z_x'^{2} + 2\,\bm p^2 Y_x^{2} + 4\,d V' Z_x' + 2\,d V' Y_x - 4\,\bm p^2 Z_x' Y_x + d \bm p^2 Y_x^{2} + 2\,d \bm p^2 Z_x' Y_x
\notag\\
&\qquad\quad + 2\,d \omega_n^{2} Z_x' Y_\tau + d \omega_n^{2} Y_x Y_\tau\Big) - \rho\,G_{LT}^{2}\Big(4\,\bm p^2 Z_x'^{2} - 4\,d V' Z_x' - 2\,d V' Y_x - 4\,\bm p^2 Z_x' Y_x - d \bm p^2 Y_x^{2}
\notag\\
&\qquad\quad - 2\,d \bm p^2 Z_x' Y_x - 2\,d \omega_n^{2} Z_x' Y_\tau - d \omega_n^{2} Y_x Y_\tau\Big)
\notag\\
&\qquad + 2 \rho \bm p^2\,G_{LL} \partial_{\bm p^2}G_{TT}\big(2\,Z_x' - Y_x\big)\big(2\,V' + \bm p^2 Y_x + \omega_n^{2} Y_\tau\big)
\notag\\
&\qquad - 2 \rho \bm p^2\,\partial_{\bm p^2}G_{LL} G_{TT}\big(2\,Z_x' - Y_x\big)\big(2\,V' + \bm p^2 Y_x + \omega_n^{2} Y_\tau\big)
\notag\\
&\qquad - 2 \rho \bm p^2\,\big(\partial_{\bm p^2}G_{LT}\big)^{2}\big(2\,V' + \bm p^2 Y_x + \omega_n^{2} Y_\tau\big)^{2}
\notag\\
&\qquad - 2 \rho \bm p^2\,\partial_{\bm p^2}G_{LL} \partial_{\bm p^2}G_{TT}\big(2\,V' + \bm p^2 Y_x + \omega_n^{2} Y_\tau\big)^{2}\Big\},
\label{eq:flot_ZxT_DE2}
\end{align}

\begin{align}
&\partial_t Z_{\tau,k}(\rho) ={} - \frac{1}{2}\,\tilde\partial_t\,\frac{1}{\beta}\sum_{\omega_n}\int_{\bm p}\,\Big\{ G_{LL}\Big(Z_\tau' + Y_\tau + 2\,\rho Z_\tau'' + 5\,\rho Y_\tau' + 2\,\rho^{2} Y_\tau''\Big) + G_{TT}\big(Z_\tau' + \rho Y_\tau'\big)
\notag\\
&\qquad + 2 \rho\,G_{LL}^{2}\big(Z_\tau' + Y_\tau + \rho Y_\tau'\big)\Big(6\,V' + 4\,\rho V'' + 2\,\bm p^2 Z_x' + 2\,\bm p^2 Y_x + 3\,\omega_n^{2} Z_\tau' + 3\,\omega_n^{2} Y_\tau + 2\,\rho \bm p^2 Y_x'
\notag\\
&\qquad\quad + 3\,\rho \omega_n^{2} Y_\tau'\Big) - 12 \rho \omega_n\,G_{LL} G_{LT}\big(2\,Z_{\tau1}' + \rho Z_{\tau1}''\big)\big(Z_\tau' + Y_\tau + \rho Y_\tau'\big)
\notag\\
&\qquad - 2 \rho\,G_{LL} G_{TT}\big(2\,Z_{\tau1}' + \rho Z_{\tau1}''\big)^{2} - 2 \rho\,G_{LT}^{2}\Big(2\,V' Z_\tau' + 5\,V' Y_\tau + 2\,\rho V' Y_\tau' + 2\,\rho V'' Y_\tau
\notag\\
&\qquad\quad + 2\,\bm p^2 Z_x' Z_\tau' + 3\,\bm p^2 Z_x' Y_\tau + \bm p^2 Y_x Y_\tau + 4\,\omega_n^{2} Z_\tau'^{2} + \omega_n^{2} Y_\tau^{2} + 5\,\omega_n^{2} Z_\tau' Y_\tau + 2\,\rho \bm p^2 Z_x' Y_\tau'
\notag\\
&\qquad\quad + \rho \bm p^2 Y_x' Y_\tau + 4\,\rho \omega_n^{2} Z_\tau' Y_\tau' + \rho \omega_n^{2} Y_\tau Y_\tau'\Big) - 4 \rho \omega_n\,G_{LT} G_{TT}\big(Z_\tau' + Y_\tau\big)\big(2\,Z_{\tau1}' + \rho Z_{\tau1}''\big)
\notag\\
&\qquad + 2 \rho\,G_{TT}^{2}\Big(V' Y_\tau + \bm p^2 Z_x' Y_\tau + \omega_n^{2} Z_\tau'^{2} + \omega_n^{2} Z_\tau' Y_\tau\Big)
\notag\\
&\qquad + 4 \rho \omega_n\,G_{LL} \partial_{\omega_n}G_{LL}\big(Z_\tau' + Y_\tau + \rho Y_\tau'\big)\Big(3\,V' + 2\,\rho V'' + \bm p^2 Z_x' + \bm p^2 Y_x + \omega_n^{2} Z_\tau' + \omega_n^{2} Y_\tau
\notag\\
&\qquad\quad + \rho \bm p^2 Y_x' + \rho \omega_n^{2} Y_\tau'\Big) + \rho\,G_{LL} \partial^2_{\omega_n}G_{LL}\Big(3\,V' + 2\,\rho V'' + \bm p^2 Z_x' + \bm p^2 Y_x + \omega_n^{2} Z_\tau' + \omega_n^{2} Y_\tau
\notag\\
&\qquad\quad + \rho \bm p^2 Y_x' + \rho \omega_n^{2} Y_\tau'\Big)^{2} - 4 \rho\,G_{LL} \partial_{\omega_n}G_{LT}\big(2\,Z_{\tau1}' + \rho Z_{\tau1}''\big)\Big(3\,V' + 2\,\rho V'' + \bm p^2 Z_x' + \bm p^2 Y_x
\notag\\
&\qquad\quad + 2\,\omega_n^{2} Z_\tau' + 2\,\omega_n^{2} Y_\tau + \rho \bm p^2 Y_x' + 2\,\rho \omega_n^{2} Y_\tau'\Big)
\notag\\
&\qquad - 2 \rho \omega_n\,G_{LL} \partial^2_{\omega_n}G_{LT}\big(2\,Z_{\tau1}' + \rho Z_{\tau1}''\big)\Big(3\,V' + 2\,\rho V'' + \bm p^2 Z_x' + \bm p^2 Y_x + \omega_n^{2} Z_\tau' + \omega_n^{2} Y_\tau
\notag\\
&\qquad\quad + \rho \bm p^2 Y_x' + \rho \omega_n^{2} Y_\tau'\Big) - 4 \rho \omega_n\,G_{LL} \partial_{\omega_n}G_{TT}\big(2\,Z_{\tau1}' + \rho Z_{\tau1}''\big)^{2}
\notag\\
&\qquad - \rho \omega_n^{2}\,G_{LL} \partial^2_{\omega_n}G_{TT}\big(2\,Z_{\tau1}' + \rho Z_{\tau1}''\big)^{2} - 4 \rho \omega_n^{2}\,\partial_{\omega_n}G_{LL} G_{LT}\big(2\,Z_{\tau1}' + \rho Z_{\tau1}''\big)\big(Z_\tau' + Y_\tau + \rho Y_\tau'\big)
\notag\\
&\qquad - 2 \rho \omega_n\,\partial^2_{\omega_n}G_{LL} G_{LT}\big(2\,Z_{\tau1}' + \rho Z_{\tau1}''\big)\Big(3\,V' + 2\,\rho V'' + \bm p^2 Z_x' + \bm p^2 Y_x + \omega_n^{2} Z_\tau' + \omega_n^{2} Y_\tau
\notag\\
&\qquad\quad + \rho \bm p^2 Y_x' + \rho \omega_n^{2} Y_\tau'\Big) - \rho \omega_n^{2}\,\partial^2_{\omega_n}G_{LL} G_{TT}\big(2\,Z_{\tau1}' + \rho Z_{\tau1}''\big)^{2}
\notag\\
&\qquad + 4 \rho \omega_n\,G_{LT} \partial_{\omega_n}G_{LT}\Big(4\,Z_{\tau1}'^{2} - 4\,V' Z_\tau' - V' Y_\tau - \rho V' Y_\tau' - 2\,\rho V'' Z_\tau' + 4\,\rho Z_{\tau1}' Z_{\tau1}'' - 2\,\bm p^2 Z_x' Z_\tau'
\notag\\
&\qquad\quad - \bm p^2 Z_x' Y_\tau - \bm p^2 Y_x Z_\tau' + \rho^{2} Z_{\tau1}''^{2} - 2\,\omega_n^{2} Z_\tau'^{2} - 2\,\omega_n^{2} Z_\tau' Y_\tau - \rho \bm p^2 Z_x' Y_\tau' - \rho \bm p^2 Y_x' Z_\tau'
\notag\\
&\qquad\quad - 2\,\rho \omega_n^{2} Z_\tau' Y_\tau'\Big) - 2 \rho\,G_{LT} \partial^2_{\omega_n}G_{LT}\Big(3\,V'^{2} + 2\,\rho V' V'' + 4\,\bm p^2 V' Z_x' + \bm p^2 V' Y_x + \bm p^{4} Z_x'^{2}
\notag\\
&\qquad\quad - 4\,\omega_n^{2} Z_{\tau1}'^{2} + \bm p^{4} Z_x' Y_x + 4\,\omega_n^{2} V' Z_\tau' + \omega_n^{2} V' Y_\tau + \rho \bm p^2 V' Y_x' + 2\,\rho \bm p^2 V'' Z_x' + \rho \bm p^{4} Z_x' Y_x'
\notag\\
&\qquad\quad + \rho \omega_n^{2} V' Y_\tau' + 2\,\rho \omega_n^{2} V'' Z_\tau' - 4\,\rho \omega_n^{2} Z_{\tau1}' Z_{\tau1}'' + 2\,\bm p^2 \omega_n^{2} Z_x' Z_\tau' + \bm p^2 \omega_n^{2} Z_x' Y_\tau + \bm p^2 \omega_n^{2} Y_x Z_\tau'
\notag\\
&\qquad\quad + \omega_n^{4} Z_\tau'^{2} - \rho^{2} \omega_n^{2} Z_{\tau1}''^{2} + \omega_n^{4} Z_\tau' Y_\tau + \rho \bm p^2 \omega_n^{2} Z_x' Y_\tau' + \rho \bm p^2 \omega_n^{2} Y_x' Z_\tau' + \rho \omega_n^{4} Z_\tau' Y_\tau'\Big)
\notag\\
&\qquad - 4 \rho\,G_{LT} \partial_{\omega_n}G_{TT}\big(2\,Z_{\tau1}' + \rho Z_{\tau1}''\big)\big(V' + \bm p^2 Z_x' + 2\,\omega_n^{2} Z_\tau'\big)
\notag\\
&\qquad - 2 \rho \omega_n\,G_{LT} \partial^2_{\omega_n}G_{TT}\big(2\,Z_{\tau1}' + \rho Z_{\tau1}''\big)\big(V' + \bm p^2 Z_x' + \omega_n^{2} Z_\tau'\big)
\notag\\
&\qquad - 4 \rho \omega_n^{2} Z_\tau'\,\partial_{\omega_n}G_{LT} G_{TT}\big(2\,Z_{\tau1}' + \rho Z_{\tau1}''\big)
\notag\\
&\qquad - 2 \rho \omega_n\,\partial^2_{\omega_n}G_{LT} G_{TT}\big(2\,Z_{\tau1}' + \rho Z_{\tau1}''\big)\big(V' + \bm p^2 Z_x' + \omega_n^{2} Z_\tau'\big)
\notag\\
&\qquad + 4 \rho \omega_n Z_\tau'\,G_{TT} \partial_{\omega_n}G_{TT}\big(V' + \bm p^2 Z_x' + \omega_n^{2} Z_\tau'\big) + \rho\,G_{TT} \partial^2_{\omega_n}G_{TT}\big(V' + \bm p^2 Z_x' + \omega_n^{2} Z_\tau'\big)^{2}\Big\},
\label{eq:flot_Ztau_DE2}
\end{align}

\begin{align}
&\partial_t \big[Z_{\tau,k}(\rho)+\rho\,Y_{\tau,k}(\rho)\big] ={} - \frac{1}{8}\,\tilde\partial_t\,\frac{1}{\beta}\sum_{\omega_n}\int_{\bm p}\,\Big\{ 4\,G_{LL}\big(Z_\tau' + 2\,\rho Z_\tau''\big) + 4\,G_{TT}\big(Z_\tau' + Y_\tau\big)
\notag\\
&\qquad - 8 \rho\,G_{LL}^{2}\big(2\,Z_{\tau1}' + \rho Z_{\tau1}''\big)^{2} - 16 \rho \omega_n\,G_{LL} G_{LT}\big(2\,Z_\tau' - Y_\tau\big)\big(2\,Z_{\tau1}' + \rho Z_{\tau1}''\big)
\notag\\
&\qquad + 4 \rho\,G_{LL} G_{TT}\Big(4\,V' Z_\tau' + 2\,V' Y_\tau + 2\,\bm p^2 Y_x Z_\tau' + \bm p^2 Y_x Y_\tau + 4\,\omega_n^{2} Z_\tau'^{2} + 3\,\omega_n^{2} Y_\tau^{2} - 2\,\omega_n^{2} Z_\tau' Y_\tau\Big)
\notag\\
&\qquad + 4 \rho\,G_{LT}^{2}\Big(4\,V' Z_\tau' + 2\,V' Y_\tau + 2\,\bm p^2 Y_x Z_\tau' + \bm p^2 Y_x Y_\tau - 4\,\omega_n^{2} Z_\tau'^{2} + \omega_n^{2} Y_\tau^{2} + 6\,\omega_n^{2} Z_\tau' Y_\tau\Big)
\notag\\
&\qquad - 8 \rho\,G_{LL} \partial_{\omega_n}G_{LT}\big(2\,Z_{\tau1}' + \rho Z_{\tau1}''\big)\big(2\,V' + \bm p^2 Y_x + \omega_n^{2} Y_\tau\big)
\notag\\
&\qquad + 8 \rho \omega_n Z_\tau'\,G_{LL} \partial_{\omega_n}G_{TT}\big(2\,V' + \bm p^2 Y_x + \omega_n^{2} Y_\tau\big) + \rho\,G_{LL} \partial^2_{\omega_n}G_{TT}\big(2\,V' + \bm p^2 Y_x + \omega_n^{2} Y_\tau\big)^{2}
\notag\\
&\qquad + 8 \rho\,\partial_{\omega_n}G_{LL} G_{LT}\big(2\,Z_{\tau1}' + \rho Z_{\tau1}''\big)\big(2\,V' + \bm p^2 Y_x + \omega_n^{2} Y_\tau\big)
\notag\\
&\qquad - 8 \rho \omega_n\,\partial_{\omega_n}G_{LL} G_{TT}\big(Z_\tau' - Y_\tau\big)\big(2\,V' + \bm p^2 Y_x + \omega_n^{2} Y_\tau\big)
\notag\\
&\qquad + \rho\,\partial^2_{\omega_n}G_{LL} G_{TT}\big(2\,V' + \bm p^2 Y_x + \omega_n^{2} Y_\tau\big)^{2} + 8 \rho \omega_n Y_\tau\,G_{LT} \partial_{\omega_n}G_{LT}\big(2\,V' + \bm p^2 Y_x + \omega_n^{2} Y_\tau\big)
\notag\\
&\qquad + 2 \rho\,G_{LT} \partial^2_{\omega_n}G_{LT}\big(2\,V' + \bm p^2 Y_x + \omega_n^{2} Y_\tau\big)^{2}\Big\},
\label{eq:flot_ZtauT_DE2}
\end{align}

\begin{align}
&\partial_t \big[Z_{\tau1,k}(\rho)+\rho\,Z_{\tau1,k}'(\rho)\big] ={} \frac{1}{2}\,\tilde\partial_t\,\frac{1}{\beta}\sum_{\omega_n}\int_{\bm p}\,\Big\{ - G_{LL}\big(2\,Z_{\tau1}' + 7\,\rho Z_{\tau1}'' + 2\,\rho^{2} Z_{\tau1}'''\big)
\notag\\
&\qquad + 2 \omega_n\,G_{LT}\Big(2\,Z_\tau' - Y_\tau + 2\,\rho Z_\tau'' - \rho Y_\tau'\Big) - G_{TT}\big(2\,Z_{\tau1}' + \rho Z_{\tau1}''\big)
\notag\\
&\qquad - 2 \rho\,G_{LL}^{2}\big(2\,Z_{\tau1}' + \rho Z_{\tau1}''\big)\Big(3\,V' + 2\,\rho V'' + \bm p^2 Z_x' + \bm p^2 Y_x + \omega_n^{2} Z_\tau' + \omega_n^{2} Y_\tau + \rho \bm p^2 Y_x' + \rho \omega_n^{2} Y_\tau'\Big)
\notag\\
&\qquad + 2 \rho \omega_n\,G_{LL} G_{LT}\Big(8\,Z_{\tau1}'^{2} - 6\,V' Z_\tau' + 3\,V' Y_\tau - 4\,\rho V'' Z_\tau' + 2\,\rho V'' Y_\tau + 8\,\rho Z_{\tau1}' Z_{\tau1}'' - 2\,\bm p^2 Z_x' Z_\tau'
\notag\\
&\qquad\quad + \bm p^2 Z_x' Y_\tau - 2\,\bm p^2 Y_x Z_\tau' + \bm p^2 Y_x Y_\tau + 2\,\rho^{2} Z_{\tau1}''^{2} - 2\,\omega_n^{2} Z_\tau'^{2} + \omega_n^{2} Y_\tau^{2} - \omega_n^{2} Z_\tau' Y_\tau - 2\,\rho \bm p^2 Y_x' Z_\tau'
\notag\\
&\qquad\quad + \rho \bm p^2 Y_x' Y_\tau - 2\,\rho \omega_n^{2} Z_\tau' Y_\tau' + \rho \omega_n^{2} Y_\tau Y_\tau'\Big)
\notag\\
&\qquad - \rho\,G_{LL} G_{TT}\big(2\,Z_{\tau1}' + \rho Z_{\tau1}''\big)\Big(2\,V' + \bm p^2 Y_x + 4\,\omega_n^{2} Z_\tau' - \omega_n^{2} Y_\tau\Big)
\notag\\
&\qquad + \rho\,G_{LT}^{2}\big(2\,Z_{\tau1}' + \rho Z_{\tau1}''\big)\Big(2\,\bm p^2 Z_x' - \bm p^2 Y_x + 6\,\omega_n^{2} Z_\tau' - 3\,\omega_n^{2} Y_\tau\Big)
\notag\\
&\qquad - 2 \rho \omega_n\,G_{LT} G_{TT}\big(2\,Z_\tau' - Y_\tau\big)\big(V' + \bm p^2 Z_x' + \omega_n^{2} Z_\tau'\big)
\notag\\
&\qquad - \rho\,G_{LL} \partial_{\omega_n}G_{LT}\big(2\,V' + \bm p^2 Y_x + \omega_n^{2} Y_\tau\big)\Big(3\,V' + 2\,\rho V'' + \bm p^2 Z_x' + \bm p^2 Y_x + \omega_n^{2} Z_\tau' + \omega_n^{2} Y_\tau
\notag\\
&\qquad\quad + \rho \bm p^2 Y_x' + \rho \omega_n^{2} Y_\tau'\Big) - \rho \omega_n\,G_{LL} \partial_{\omega_n}G_{TT}\big(2\,Z_{\tau1}' + \rho Z_{\tau1}''\big)\big(2\,V' + \bm p^2 Y_x + \omega_n^{2} Y_\tau\big)
\notag\\
&\qquad + \rho\,\partial_{\omega_n}G_{LL} G_{LT}\big(2\,V' + \bm p^2 Y_x + \omega_n^{2} Y_\tau\big)\Big(3\,V' + 2\,\rho V'' + \bm p^2 Z_x' + \bm p^2 Y_x + \omega_n^{2} Z_\tau' + \omega_n^{2} Y_\tau
\notag\\
&\qquad\quad + \rho \bm p^2 Y_x' + \rho \omega_n^{2} Y_\tau'\Big) + \rho \omega_n\,\partial_{\omega_n}G_{LL} G_{TT}\big(2\,Z_{\tau1}' + \rho Z_{\tau1}''\big)\big(2\,V' + \bm p^2 Y_x + \omega_n^{2} Y_\tau\big)
\notag\\
&\qquad - \rho\,G_{LT} \partial_{\omega_n}G_{TT}\big(V' + \bm p^2 Z_x' + \omega_n^{2} Z_\tau'\big)\big(2\,V' + \bm p^2 Y_x + \omega_n^{2} Y_\tau\big)
\notag\\
&\qquad + \rho\,\partial_{\omega_n}G_{LT} G_{TT}\big(V' + \bm p^2 Z_x' + \omega_n^{2} Z_\tau'\big)\big(2\,V' + \bm p^2 Y_x + \omega_n^{2} Y_\tau\big)\Big\},
\label{eq:flot_Ztau1_DE2}
\end{align}
where the propagators, obtained from the
ansatz~\eqref{eq:ansatz_DE2_annexe} according to
Eq.~\eqref{eq:Gk_definition}, read
\begin{equation}
\begin{aligned}
G_{LL} &= -\frac{R_k + Z_x\,\bm p^2 + Z_\tau\,\omega_n^2 + V}{D}, \\
G_{TT} &= -\frac{R_k + \big(Z_x+\rho\,Y_x\big)\bm p^2
+ \big(Z_\tau+\rho\,Y_\tau\big)\omega_n^2 + V + 2\rho\,V'}{D}, \\
G_{LT} &= \frac{\big(Z_{\tau1}+\rho\,Z_{\tau1}'\big)\,\omega_n}{D}, \\
D &= \big(R_k + Z_x\,\bm p^2 + Z_\tau\,\omega_n^2 + V\big)
\big(R_k + (Z_x+\rho\,Y_x)\bm p^2 + (Z_\tau+\rho\,Y_\tau)\omega_n^2
+ V + 2\rho\,V'\big) \\
&\quad + \big(Z_{\tau1}+\rho\,Z_{\tau1}'\big)^2\omega_n^2 ,
\end{aligned}
\label{eq:propagateurs_DE2_annexe}
\end{equation}
with $R_k\equiv R_k(\bm p)$.

For convenience, we have given the flows of the combinations $Z_{x,k}+\rho\,Y_{x,k}$ and $Z_{\tau,k}+\rho\,Y_{\tau,k}$, which arise from projections onto the longitudinal component
$\Gamma_{k,A}+2\rho\,\Gamma_{k,B}$ of the two-point vertex. The flows of $Y_{x,k}$ and $Y_{\tau,k}$ follow from
\begin{equation}
\begin{aligned}
\partial_t Y_{x,k}
&= \frac{\partial_t\big(Z_{x,k}+\rho\,Y_{x,k}\big) - \partial_t Z_{x,k}}{\rho}, \\
\partial_t Y_{\tau,k}
&= \frac{\partial_t\big(Z_{\tau,k}+\rho\,Y_{\tau,k}\big) - \partial_t Z_{\tau,k}}{\rho}.
\end{aligned}
\label{eq:flot_Y_extraction}
\end{equation}
Similarly, projecting the flow of $\Gamma_{k,C}$ onto its term linear in $\omega_n$, Eq.~\eqref{eq:flot_Ztau1}, yields the flow of the combination $Z_{\tau1,k}+\rho\,Z_{\tau1,k}'$. The flow of $Z_{\tau1,k}$ itself can be reconstructed by extracting that of $Z_{\tau1,k}'$ from the three-point vertex $\Gamma^{(3)}_{k,212}$.

The flow equations~\eqref{eq:flot_V_DE2}--\eqref{eq:flot_Ztau1_DE2} reduce to those of the classical $O(2)$ model at DE$_2$ when only the static mode $\omega_n=0$ is retained in the Matsubara sum. This amounts to replacing $\frac{1}{\beta}\sum_{\omega_n}$ by $T$ and setting $Z_\tau=Y_\tau=Z_{\tau1}=0$. In this case, $G_{LT}=0$ and the denominator $D$ factorizes, so that
\begin{equation}
\begin{aligned}
G_{LL} &= -\frac{1}{R_k + \big(Z_x+\rho\,Y_x\big)\bm p^2 + V + 2\rho\,V'}, \\
G_{TT} &= -\frac{1}{R_k + Z_x\,\bm p^2 + V},
\end{aligned}
\end{equation}
which are respectively the longitudinal and transverse propagators of the $O(2)$ model at DE$_2$. The flows of $V_k$, $Z_{x,k}$ and $Y_{x,k}$, Eqs.~\eqref{eq:flot_V_DE2}--\eqref{eq:flot_ZxT_DE2}, then reproduce the flow equations of the classical $O(2)$ model in dimension $d$ at DE$_2$ given for instance in Ref.~\cite{jakubczyk_2014}, up to the temperature $T$ that can be absorbed into a redefinition of the field and of the potential.

The LPA$'\times$DE$_2$ flow equations are obtained from the DE$_2$ equations by treating the quantum sector at the LPA$'$ level, i.e., by setting $Y_\tau=0$ and replacing $Z_{\tau1}(\rho)$ by its value at the minimum of the effective potential, $Z_{\tau1,k}\equiv Z_{\tau1,k}(\rho_{0,k})$, its field derivatives being set to zero, and neglecting the second-order time-derivative term, $Z_\tau=0$. The flow equations then reduce to
\begin{align}
&\partial_t V_k(\rho) ={} - \frac{1}{2}\,\tilde\partial_t\,\frac{1}{\beta}\sum_{\omega_n}\int_{\bm p}\,\Big\{ G_{LL}\Big(V' + 2\,\rho V'' + \bm p^2 Z_x' + \rho \bm p^2 Y_x'\Big)
\notag\\
&\qquad + G_{TT}\big(3\,V' + \bm p^2 Z_x' + \bm p^2 Y_x\big) + \rho\,G_{LL} G_{TT}\big(2\,V' + \bm p^2 Y_x\big)^{2} + \rho\,G_{LT}^{2}\big(2\,V' + \bm p^2 Y_x\big)^{2}\Big\},
\label{eq:flot_V_LPAp}
\end{align}

\begin{align}
&\partial_t Z_{x,k}(\rho) ={} -\frac{1}{2d}\,\tilde\partial_t\,\frac{1}{\beta}\sum_{\omega_n}\int_{\bm p}\,\Big\{ d\,G_{LL}\Big(Z_x' + Y_x + 2\,\rho Z_x'' + 5\,\rho Y_x' + 2\,\rho^{2} Y_x''\Big)
\notag\\
&\qquad + d\,G_{TT}\big(Z_x' + \rho Y_x'\big) + 2 \rho\,G_{LL}^{2}\big(Z_x' + Y_x + \rho Y_x'\big)\Big(6\,d V' + \bm p^2 Z_x' + \bm p^2 Y_x + 4\,d \rho V''
\notag\\
&\qquad\quad + 2\,d \bm p^2 Z_x' + 2\,d \bm p^2 Y_x + \rho \bm p^2 Y_x' + 2\,d \rho \bm p^2 Y_x'\Big)
\notag\\
&\qquad - 2 \rho\,G_{LT}^{2}\Big(2\,\bm p^2 Z_x'^{2} + 2\,d V' Z_x' + 5\,d V' Y_x + 2\,\bm p^2 Z_x' Y_x + 2\,d \bm p^2 Z_x'^{2} + d \bm p^2 Y_x^{2} + 2\,d \rho V' Y_x'
\notag\\
&\qquad\quad + 2\,d \rho V'' Y_x + 3\,d \bm p^2 Z_x' Y_x + 2\,\rho \bm p^2 Z_x' Y_x' + 2\,d \rho \bm p^2 Z_x' Y_x' + d \rho \bm p^2 Y_x Y_x'\Big)
\notag\\
&\qquad + 2 \rho\,G_{TT}^{2}\big(\bm p^2 Z_x'^{2} + d V' Y_x + d \bm p^2 Z_x' Y_x\big) - 4 \rho \bm p^2\,\big(\partial_{\bm p^2}G_{LL}\big)^{2}\Big(3\,V' + 2\,\rho V'' + \bm p^2 Z_x' + \bm p^2 Y_x
\notag\\
&\qquad\quad + \rho \bm p^2 Y_x'\Big)^{2} + 8 \rho \bm p^2\,\big(\partial_{\bm p^2}G_{LT}\big)^{2}\big(V' + \bm p^2 Z_x'\big)\Big(3\,V' + 2\,\rho V'' + \bm p^2 Z_x' + \bm p^2 Y_x + \rho \bm p^2 Y_x'\Big)
\notag\\
&\qquad - 4 \rho \bm p^2\,\big(\partial_{\bm p^2}G_{TT}\big)^{2}\big(V' + \bm p^2 Z_x'\big)^{2}\Big\},
\label{eq:flot_Zx_LPAp}
\end{align}

\begin{align}
&\partial_t \big[Z_{x,k}(\rho)+\rho\,Y_{x,k}(\rho)\big] ={} -\frac{1}{2d}\,\tilde\partial_t\,\frac{1}{\beta}\sum_{\omega_n}\int_{\bm p}\,\Big\{ d\,G_{LL}\big(Z_x' + 2\,\rho Z_x''\big) + d\,G_{TT}\big(Z_x' + Y_x\big)
\notag\\
&\qquad + \rho\,G_{LL} G_{TT}\Big(4\,\bm p^2 Z_x'^{2} + 2\,\bm p^2 Y_x^{2} + 4\,d V' Z_x' + 2\,d V' Y_x - 4\,\bm p^2 Z_x' Y_x + d \bm p^2 Y_x^{2} + 2\,d \bm p^2 Z_x' Y_x\Big)
\notag\\
&\qquad - \rho\,G_{LT}^{2}\Big(4\,\bm p^2 Z_x'^{2} - 4\,d V' Z_x' - 2\,d V' Y_x - 4\,\bm p^2 Z_x' Y_x - d \bm p^2 Y_x^{2} - 2\,d \bm p^2 Z_x' Y_x\Big)
\notag\\
&\qquad + 2 \rho \bm p^2\,G_{LL} \partial_{\bm p^2}G_{TT}\big(2\,Z_x' - Y_x\big)\big(2\,V' + \bm p^2 Y_x\big)
\notag\\
&\qquad - 2 \rho \bm p^2\,\partial_{\bm p^2}G_{LL} G_{TT}\big(2\,Z_x' - Y_x\big)\big(2\,V' + \bm p^2 Y_x\big) - 2 \rho \bm p^2\,\big(\partial_{\bm p^2}G_{LT}\big)^{2}\big(2\,V' + \bm p^2 Y_x\big)^{2}
\notag\\
&\qquad - 2 \rho \bm p^2\,\partial_{\bm p^2}G_{LL} \partial_{\bm p^2}G_{TT}\big(2\,V' + \bm p^2 Y_x\big)^{2}\Big\},
\label{eq:flot_ZxT_LPAp}
\end{align}

\begin{align}
&\partial_t Z_{\tau1,k} ={} - \frac{1}{2}\,\tilde\partial_t\,\frac{1}{\beta}\sum_{\omega_n}\int_{\bm p}\,\Big\{ \rho\,G_{LL} \partial_{\omega_n}G_{LT}\big(2\,V' + \bm p^2 Y_x\big)\Big(3\,V' + 2\,\rho V'' + \bm p^2 Z_x' + \bm p^2 Y_x
\notag\\
&\qquad\quad + \rho \bm p^2 Y_x'\Big) - \rho\,\partial_{\omega_n}G_{LL} G_{LT}\big(2\,V' + \bm p^2 Y_x\big)\Big(3\,V' + 2\,\rho V'' + \bm p^2 Z_x' + \bm p^2 Y_x + \rho \bm p^2 Y_x'\Big)
\notag\\
&\qquad + \rho\,G_{LT} \partial_{\omega_n}G_{TT}\big(V' + \bm p^2 Z_x'\big)\big(2\,V' + \bm p^2 Y_x\big) - \rho\,\partial_{\omega_n}G_{LT} G_{TT}\big(V' + \bm p^2 Z_x'\big)\big(2\,V' + \bm p^2 Y_x\big)\Big\},
\label{eq:flot_Ztau1_LPAp}
\end{align}
where the flow of $Z_{\tau1,k}$ is evaluated at the minimum of the effective potential, $\rho=\rho_{0,k}$, in accordance with Eq.~\eqref{eq:flot_Ztau1}. 
Finally, the field-truncated LPA$'$ is obtained by additionally setting $Y_x=0$, replacing $Z_{x}(\rho)$ by its value at the minimum $Z_{x,k}(\rho_{0,k})$, and truncating the effective potential at second order in $\rho-\rho_{0,k}$. The corresponding flow equations are given in Ref.~\cite{dupuis_2009a}. Here too, we used Wetterich's regulator Eq.~\eqref{eq:regulateur_exp} with $\alpha=2.1$.

\bibliography{bibli}

\end{document}